\documentclass[aip,reprint]{revtex4-1}
\usepackage{graphicx}
\usepackage{multirow}
\usepackage{amsmath}
\usepackage{amsfonts}
\usepackage{xcolor}
\draft 

\begin{document}

\title{State preparation and evolution in quantum computing: a perspective from Hamiltonian moments}
\author{Joseph C. Aulicino}
\address{Pritzker Molecular Engineering, University of Chicago, 5640 S Ellis Ave, Chicago, IL 60637, United States of America}
\author{Trevor Keen}
\address{Department of Physics and Astronomy, University of Tennessee, Knoxville, Tennessee 377996, United States of America}
\author{Bo Peng}
\email{peng398@pnnl.gov} 
\address{Physical Sciences and Computational Division, Pacific Northwest National Laboratory, Richland, WA 99354, United States of America}

\begin{abstract}
Quantum algorithms on noisy intermediate-scale quantum (NISQ) devices are expected to soon have the ability to simulate quantum systems 
that are classically intractable, demonstrating a quantum advantage. However, the non-negligible gate error present on NISQ devices 
impedes the implementation of conventional quantum algorithms. Practical strategies usually exploit hybrid quantum-classical algorithms 
to demonstrate potentially useful applications of quantum computing in the NISQ era. Among the numerous hybrid quantum-classical algorithms, 
recent efforts highlight the development of quantum algorithms based upon quantum computed Hamiltonian moments, 
$\langle \phi | \hat{\mathcal{H}}^n | \phi \rangle$ ($n=1,2,\cdots$), with respect to quantum state $|\phi\rangle$. In this tutorial review, 
we will give a brief review of these quantum algorithms with focuses on the typical ways of computing Hamiltonian moments 
using quantum hardware and improving the accuracy of the estimated state energies based on the quantum computed moments. 
Furthermore, we will present an example to show how we can measure and compute the Hamiltonian moments of a four-site Heisenberg model, 
and compute the energy and magnetization of the model utilizing the imaginary time evolution on the real IBM-Q NISQ hardware. 
Along this line, we will further discuss some practical issues associated with these algorithms. We will conclude this tutorial review by 
discussing some possible developments and applications in this direction in the near future. 
\end{abstract}


\maketitle

\tableofcontents{}
\newpage

\section{Introduction}

The original idea of solving many-body quantum mechanical problems on a quantum computer dates back to R. Feynman's insight almost four 
decades ago \cite{Feynman1982}, when high-performance classical computing was at its early state and quantum computing was only a thought 
experiment. In the last forty years, developments in classical quantum approaches and high performance computing technology yielded adequate 
tools to perform highly accurate small- to medium-size quantum simulations and qualitative large-scale quantum simulations. Despite these 
exciting developments and achievements, the classical quantum approach is gradually approaching the computing limit attributed to the 
unfavorable scaling along with the expanded system size. For instance, consider the well-known examples of full Configuration Interaction 
(full CI) approach \cite{Sherrill1999} and Coupled Cluster (CC) approach \cite{Bartlett2007}. Classically, in order to achieve the chemical 
accuracy ($\sim$1 kcal/mol) for the energy evalulation, the runtime in the full CI and the ``golden standard" CCSD(T) scale as 
$\mathcal{O}\Big(\substack{N\\n}\Big)$ and $\mathcal{O}(N^7)$, respectively, with respect to the system size $N$ 
(and number of electrons $n$), which impedes their routine applications for large systems. On the other hand, quantum computing
(simulation performed on a quantum computer) is a next-generation computing technology that bears the hope of outperforming 
high-performance classical computing techniques to solve a wide class of problems encountered in scientific research and 
industrial applications. Quantum computation enables the encoding of an exponential amount of information about many-body 
quantum systems into a polynomial number of qubits, which provides a more efficient way to expand and explore the computational 
state-space in polynomial time \cite{sugisaki2019quantum}. 

Since the first application of quantum computing algorithm for computational chemistry problems in 2005 \cite{Aspuru-Guzik2005}, numerous quantum algorithms have been proposed aiming at demonstrating the potential quantum speedup, in comparison to the classical approaches, for solving problems in domain science such as chemistry, nuclear physics, quantum field theory, and high energy physics \cite{PhysRevX.6.031007,Linke2017experimental,Kandala2017hardware,Dumitrescu2018cloud,Klco2018quantum,colless2018computation,mccaskey2019quantum}.
However, many proposed quantum algorithms featuring favorable scaling typically have deep circuit demands, and are not suitable for the near-term noisy intermediate-scale quantum (NISQ) devices \cite{preskill2018quantum}.
In the search for quantum algorithms suitable running on the NISQ devices, one should mention hybrid quantum-classical Variational Quantum algorithm (VQA) \cite{peruzzo2014variational,mcclean2016theory,romero2018strategies,shen2017quantum,Kandala2017hardware,kandala2018extending,colless2018computation, huggins2020non},
quantum approximate optimization algorithm (QAOA) \cite{farhi2014quantum},
quantum annealing \cite{bharti2021noisy,albash2018adiabatic},
gaussian boson sampling \cite{Aaronson2011},
analog quantum simulation \cite{trabesinger2012quantum,georgescu2014quantum},
iterative quantum assisted eigensolver \cite{mcardle2019variational,motta2020determining,parrish2019quantum,kyriienko2020quantum},
and many others, which bear the hope to capitalize the near-term NISQ quantum devices. 
Nevertheless, these NISQ quantum algorithms have their own limitations. For example, VQAs face challenges such as Barren plateau \cite{mcclean2018barren,cerezo2021cost,Cerezo2021higher}, ansatz searching \cite{herasymenko2019diagrammatic}, and measurement overhead \cite{bonetmonroig2020nearly} which stimulate the emerging of some many recent improvements. 
Numerous comprehensive reviews on the NISQ quantum algorithms and techniques have appeared in recent years to pave the foundation of and 
provide guidance to future studies (see e.g. Refs.  \citenum{yuan2021hybrid,bharti2021noisy} for more recent ones). However, due to the 
rapid development of this emerging field, it becomes challenging for the reviews to comprehensively cover many newly developed quantum 
algorithms and techniques in a timely manner. Furthermore, it is usually more efficient, in particular for non-specialists, to choose a 
thread to understand and digest the basic ideas and the interconnection of these algorithms by going through some tutorials. With these 
concerns in mind, instead of giving another comprehensive review covering some newly developed quantum algorithms, here we present a 
tutorial review on some recently developed quantum algorithms based on the quantum computation of Hamiltonian moments. The Hamiltonian 
moment here is referring to the expectation value of Hamiltonian powers with respect to a given state $|\phi\rangle$ for the systems of 
interest $\langle \phi | \hat{\mathcal{H}}^n | \phi \rangle$ ($n=1,2,\cdots$), and can be considered as one of the building blocks for 
performing variational or perturbative calculations solving the energy and property of a many-body quantum system. We will introduce the 
state preparation and evolution, and hybrid algorithms associated with quantum Hamiltonian moments. With the examples 
and discussions, we hope to provide a relatively easy-to-read review with a clear demonstration of the connections and differences 
between some relevant quantum algorithms.

\begin{figure*}
\centering
\includegraphics[width=\linewidth]{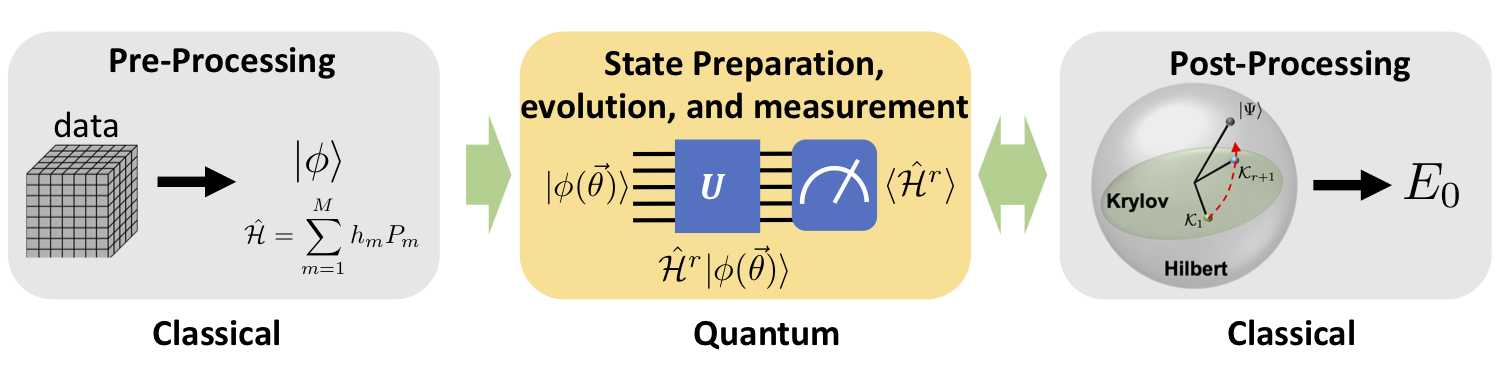}
\caption{The workflow of a typical hybrid quantum classical algorithm based on quantum computed Hamiltonian moments $\langle \hat{\mathcal{H}}^r\rangle$ ($r=1,2,\cdots$).} \label{workflow}
\end{figure*}

This tutorial review is organized as follows. Section II reviews the ways of quantum computing Hamiltonian moments. Section III reviews 
some recently proposed hybrid approaches based on quantum Hamiltonian moments (see Fig. \ref{workflow} for a typical 
workflow of the hybrid algorithm of this kind). Section IV gives a tutorial of how to target the ground state and magnetization of a 
generalized four-site Heisenberg model employing quantum Hamilonian moments in the imaginary time evolution approach on IBM-Q quantum 
hardware. Section IV also includes the discussions of some practical issues associated with performing quantum computing on the NISQ 
device with a focus on the impact of these issues on the results of quantum simulations. We will conclude this tutorial review by pointing 
out possible developments and applications in this direction in the near future. 

\section{Quantum computation of Hamiltonian moments}

Given a system Hamiltonian $\hat{\mathcal{H}}$ of interest and a trial state $|\phi\rangle$, the $n$-th order ($n\geq1$) Hamiltonian moment with respect to $|\phi\rangle$ is defined as
\begin{align}
	\langle \hat{\mathcal{H}}^n \rangle \equiv \langle \phi | \hat{\mathcal{H}}^n | \phi \rangle. \label{ham_n}
\end{align}
In classical computations, $\langle \hat{\mathcal{H}}^n \rangle$'s are obtained by power iteration, i.e. a repeated multiplication of 
$\hat{\mathcal{H}}$ and $| \phi \rangle$. However, as the dimension of the Hilbert space grows exponentially with the system size, and 
the number of terms that constitutes $\langle \hat{\mathcal{H}}^n \rangle$ would quickly blow up as the power $n$ becomes large, the 
classical computations of $\langle \hat{\mathcal{H}}^n \rangle$'s would quickly become intractable.
On the other hand, quantum computing allows for a more straightforward encoding of the quantum states defined in a Hilbert space of potentially large and classically intractable dimensions.
Therefore, we consider (\textbf{a}) how to utilize quantum resources for the direct computation of $\langle \hat{\mathcal{H}}^n \rangle$, 
and (\textbf{b}) how to exploit the noisy $\langle \hat{\mathcal{H}}^n \rangle$'s computed on NISQ devices to accurately evaluate 
ground/excited states and dynamics of quantum many-body systems. 
In this section, we try to answer the first question by briefly reviewing some of the methods that have been proposed for directly 
computing  $\langle \hat{\mathcal{H}}^n \rangle$ on quantum devices. We leave the discussion of the second question to Section III.

To directly compute $\langle \hat{\mathcal{H}}^n \rangle$ on quantum devices for a quantum many-body system, a first step is to encode the many-body Hamiltonian in terms of qubits.  There have been many ways proposed to encode the system Hamiltonian. For example, in the Jordan-Wigner \cite{JW1928} and Bravyi-Kitaev methods \cite{bravyi2002fermionic, seeley2012bravyi, tranter2015BK}, the $N_q$-qubit system Hamiltonian is encoded as a linear combination of tensor products of Pauli qubit operators,
\begin{align}
	\hat{\mathcal{H}} = \sum_{m=1}^M h_m P_m, \label{ham}
\end{align}
where $h_m$ is a scalar and $P_m\equiv{\prod_{q=1}^{N_q}}^{\otimes}\sigma_i(q)$ is a Pauli string with $\sigma_i(q)$ being either $2\times2$ identity matrix or one of the three Pauli matrices 
\begin{align}
\sigma_x = \left(\begin{array}{cc} 0 & 1 \\ 1 & 0 \end{array}\right),~~
\sigma_y = \left(\begin{array}{cc} 0 & -i \\ i & 0 \end{array}\right),~~
\sigma_z = \left(\begin{array}{cc} 1 & 0 \\ 0 & -1 \end{array}\right) \notag
\end{align}
for the $q$-th qubit. From (\ref{ham}), a naive way to directly compute $\langle \hat{\mathcal{H}}^n \rangle$ is to plug (\ref{ham}) into (\ref{ham_n}), and evaluate the expectation values of the products of Pauli strings that constitute $\langle \hat{\mathcal{H}}^n \rangle$. Apparently, without any simplification or reduction, the naive way would require evaluating $\mathcal{O}(M^n)$ terms and would quickly become intractable as $M$ and/or $n$ scale up. Therefore, major efforts in this direction focus on mitigating this evaluation overhead as much as possible by introducing some simplifications and reductions based on the properties of the Pauli strings.

\subsection{Term-by-term measurement} \label{measurement}

One straightforward way is to apply Pauli reduction and commutativity to reduce the number of terms, and then do term-by-term measurements.
The Pauli reduction simply utilizes the commutation and anti-commutation relations between Pauli matrices
\begin{align}
\left\{\begin{array}{cl}
[\sigma_i,\sigma_j] &= 2i \epsilon_{ijk}\sigma_k \\
\{\sigma_i,\sigma_j\} &= 2\delta_{ij}I
\end{array}\right. 
\Rightarrow \sigma_i\sigma_j = i \epsilon_{ijk}\sigma_k + \delta_{ij}I, \label{Pauli_reduction}
\end{align}
where $\sigma_i,\sigma_j,\sigma_k$ are Pauli matrices, $\epsilon_{ijk}$ is the structure constant following the Levi-Civita symbol \cite{tyldesley1975introduction}, and $\delta_{ij}$ is the Kronecker delta. From (\ref{Pauli_reduction}) it is straightforward to see the product of Pauli strings is another Pauli string with an appropriate phase factor. Therefore, in comparison to the evaluation of $\langle \hat{\mathcal{H}} \rangle$, the evaluation of $\langle \hat{\mathcal{H}}^n \rangle$ does not increase the circuit depth. Furthermore, we notice that the measurement of a single Pauli string could be used to determine classes of contributions to Hamiltonian moments of arbitrary order. For example, since every Pauli string is unitary, the evaluation of $\langle P_i\rangle$ can simultaneously contribute all the $\langle H^{2k+1}\rangle$'s ($k\geq 0, k\in \mathbb{N}$) \cite{kowalski2020quantum}. Suchsland et al. systematically analyzed the actual number of Pauli strings corresponding to $\hat{\mathcal{H}}^n$ ($n=1,2,3$) for a collection of 11 molecules, including H$_n$ ($n=2,4,6,8,10$), LiH, NaH, H$_2$O, NH$_3$, and N$_2$, that require up to $N_q=18$ qubits \cite{Suchsland2021algorithmicerror}. It was found that the actual numbers of Pauli strings in these systems after applying (\ref{Pauli_reduction}) exhibit at most $\mathcal{O}(N_q^3)$, $\mathcal{O}(N_q^5)$, and $\mathcal{O}(N_q^{7})$ scalings in comparison to the predicted $\mathcal{O}(N_q^4)$, $\mathcal{O}(N_q^8)$, and $\mathcal{O}(N_q^{12})$ scalings for $\hat{\mathcal{H}}^n$ ($n=1,2,3$) respectively.

\begin{figure}[h!]
\centering
\includegraphics[width=\linewidth]{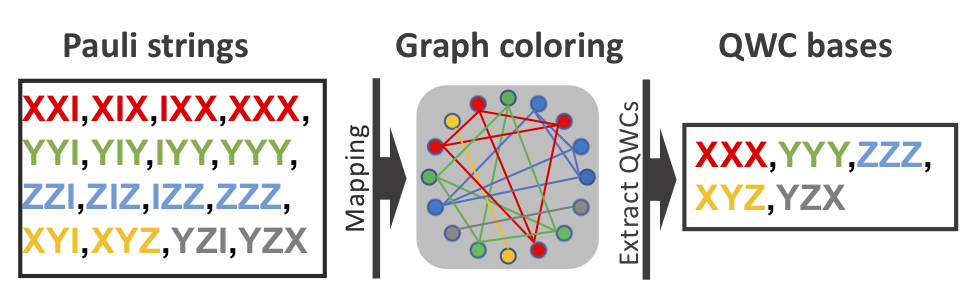}
\caption{A pictorial QWC grouping procedure.} \label{qwc}
\end{figure}

On the top of the above simplification, a more significant term reduction can be achieved by utilizing the commutativity of the Pauli strings. 
To see this method works mathematically, consider operators $\hat{\mathcal{A}}$ and $\hat{\mathcal{B}}$ that commute, i.e. 
$[\hat{\mathcal{A}},\hat{\mathcal{B}}] = 0$, then there exists a complete orthonormal eigenbasis ${|u_i\rangle}$ ($i=0,1,\cdots$) 
that simultaneously diagonalizes both operators, and we can write
\begin{align}
\hat{\mathcal{A}} = \sum_i \lambda_i |u_i\rangle \langle u_i |, ~~
\hat{\mathcal{B}} = \sum_i \kappa_i |u_i\rangle \langle u_i |,
\end{align}
with $\{\lambda_i, i=0,1,\cdots\}$ and $\{\kappa_i, i=0,1,\cdots\}$ being the eigenvalues of $\hat{\mathcal{A}}$ and $\hat{\mathcal{B}}$, respectively. Now the expectation values of $\hat{\mathcal{A}}$ and $\hat{\mathcal{B}}$ with respect to a trial state $|\phi\rangle$ can be expressed as
\begin{align}
    \langle \hat{\mathcal{A}} \rangle &= \langle \phi | \hat{\mathcal{A}} | \phi \rangle = \sum_i \lambda_i |\langle u_i|\phi\rangle|^2, \notag \\
    \langle \hat{\mathcal{B}} \rangle &= \langle \phi | \hat{\mathcal{B}} | \phi \rangle = \sum_i \kappa_i |\langle u_i|\phi\rangle|^2. \label{simultaneous}
\end{align}
Therefore, if we measure each $|\langle u_i | \phi \rangle|^2$'s, we can deduce $\langle \hat{\mathcal{A}} \rangle$ and $\langle \hat{\mathcal{B}} \rangle$ based on (\ref{simultaneous}). 

The simplest commutativity among Pauli strings is qubitwise commutativity (QWC) \cite{Kandala2017hardware,mcclean2016theory}, where two Pauli strings qubitwise commute if two single-qubit Pauli matrices at each index commute, i.e.
\begin{align}
&[I,I] = [I,\sigma_x] = [I,\sigma_y] = [I,\sigma_z] = 0, \notag \\
&[\sigma_x,\sigma_x] = [\sigma_y,\sigma_y] = [\sigma_z,\sigma_z] = 0.
\end{align}
For example, $\{\sigma_x\otimes \sigma_y, \sigma_x\otimes I, I\otimes \sigma_y, I\otimes I\}$ is a QWC class, 
since any pair of Pauli strings in this class commute (a more pictorial QWC grouping procedure is shown in Fig. \ref{qwc}). 
For the observables grouped in one QWC class, one can easily find a shared eigenbasis that simultaneously 
diagonalizes all the observables. Take the previous example, if we want to measure the QWC class in the 
computational basis, utilizing the Clifford representation of single Pauli matrices $\sigma_x$ and $\sigma_y$
\begin{align}
&\sigma_x = H \sigma_z H, \notag \\
&\sigma_y = S \sigma_x S^\dagger = S H \sigma_z H S^\dagger,
\end{align}
we will only need perform the following measurement 

\begin{figure}[h]
\centering
\includegraphics[scale=0.5]{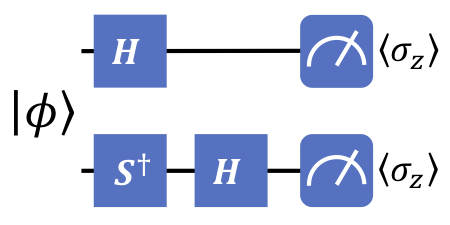}
\end{figure}
\noindent which gives the probabilities of the variational circuit rotated into the shared eigenbasis, and the expectation values of each Pauli string can then be recovered by classically combining these probabilities with the eigenvalues of these Pauli strings. 

Now, given a large amount of Pauli strings that constitute $\hat{\mathcal{H}}^n$, how should we find an optimal grouping of these Pauli strings such that the total number of measurements is minimized? Unfortunately, it turns out that finding the optimal grouping is NP-hard \cite{Karp2010,gokhale2020}, and can be converted to minimum clique cover or minimum graph coloring problem, which in practice can be approximately solved through heuristic approaches that scale quadratically with the number of the Pauli strings \cite{Izmaylov2020}. In terms of term reduction, for small molecules the QWC grouping can bring roughly 75\% saving for evaluating $\langle \hat{\mathcal{H}}\rangle$, and even bigger savings for for $\langle \hat{\mathcal{H}}^n\rangle$ ($n>1$). Early studies of applying QWC bases to evaluating $\langle \hat{\mathcal{H}}^4\rangle$ for the Heisenberg models represented by 10 to 40 qubits exhibits a sub-linear scaling of the number of measurements in the number of the qubits \cite{Vallury2020quantumcomputed}. Furthermore, for some small molecular systems, it has been shown that the number of QWC bases grouping the Pauli strings that constitute $\hat{\mathcal{H}}^n$ eventually reach a plateau, regardless of the power $n$ \cite{Claudino_2021}.

Essentially, the QWC grouping has its roots in mutually unbiased bases (MUB) \cite{Schwinger1960unitary,klappenecker2004mub} from 
quantum information theory associated with maximizing the information learned from a single measurement. In general, MUBs group the 
potentially $4^{N_q}-1$ $N_q$-qubit Pauli strings (excluding identity string) into commuting groups of maximal size, of which using 
QWCs would allow reducing this requirement to $3^{N_q}$ \cite{Altepeter2004}. Further, it is known that there in principle exists an 
MUB grouping for $N_q$-qubit that can further reduce this requirement to even $2^{N_q}+1$ with maximum $2^{N_q}-1$ distinct Pauli 
strings in each commuting group, however, the entanglement in the MUBs \cite{PhysRevA.65.032320} makes MUB quantum state tomography 
challenging. In practice, beside QWC, other grouping techniques have alternatively been proposed, including general 
commutativity \cite{gokhale2020,Izmaylov2020}, unitary partitioning \cite{Izmaylov2019}, and Fermionic basis rotation 
grouping \cite{Huggins_2021}. Generally speaking, the applications of these grouping rules for evaluating $\langle \hat{\mathcal{H}} \rangle$ 
have shown that, at a cost of introducing additional one-/multi-qubit unitary transformation before the measurement, 
the total number of terms can be significantly reduced from $\mathcal{O}(N_q^4)$ to $\mathcal{O}(N_q^{2\sim3})$. 
In particular, for simpler cases where the entire Hamiltonian or its power could even be transformed by a single unitary, 
the evaluation of the corresponding expectation values can be done in a single set of measurements. 

It is worth mentioning that the above discussion is limited to the number of terms that can be efficiently reduced through 
the groupings that feature the commutativity of Pauli strings, while the total number of measurements required from 
the number of groups also critically depends on the covariance between Pauli strings, $\text{cov}\big(P_i,P_j\big)$, 
and the desired precision $\epsilon$. Given that we can always write the Hamiltonian powers as linear combination of 
Pauli strings similar to Eq. (\ref{ham}), the total number of measurements for evaluating the moments can then be 
expressed as \cite{gonthier2020identifying,Rubin2018Hybrid,PhysRevA.92.042303}
\begin{align}
  & \text{\# of Measurements} = \notag \\
   &~~~~~~~~~~~~~~~~~~~~\Bigg( \displaystyle\frac{\sum_G\sqrt{\sum_{i,j,\in G}h_i h_j \text{cov}\big(P_i,P_j\big)}}{\epsilon}\Bigg)^2
   \label{eq:Measurement}
\end{align}
with $G$ indexing the groups. Therefore, it is likely that the number of groups decreases at the cost of introducing larger covariances that could essentially increase the total number of measurements required to achieve a desired precision. 

There are also many other advanced measurement schemes proposed recently. One example is to simultaneously obtain expectation values of multiple observables by randomly measuring and projecting the quantum state into classical shadows \cite{huang2020predicting,huang2021efficient,aaronson2018shadow,struchalin2021experimental,chen2021robust,zhao2021fermionic,acharya2021informationally,hadfield2021adaptive,hillmich2021decision,zhang2021experimental}. The algorithm in principle enables the measurements of $M$ low-weight observables using only $\mathcal{O}(\log_2M)$ samples. The practical performance of the algorithm for model and molecular Hamiltonians on NISQ device, in terms of accuracy and efficiency, is still under intense study.

\subsection{Expansion by Chebyshev polynomials via quantum walk}
Another way to exactly implement $\hat{\mathcal{H}}^n$ is by Chebyshev polynomial expansion. We can exactly express a monomial $x^n$ on $\left[-1,1\right]$ as a finite sum of Chebyshev polynomials of the first kind, i.e. 
\begin{align}
    x^n = \sum\limits_{k=0}^n C_{nk} T_k (x)
\end{align}
where
\begin{align}
C_{nk} =
\begin{cases}
    \dfrac{1}{2^{n-1}} ~\binom{n}{(n-k)/2} & \text{if }(n-k) \text{ is even, } j>0 \\[1em]
    \dfrac{1}{2^{n}} ~\binom{n}{n/2} & \text{if }n \text{ is even, } j=0 \\[1em]
    0 & \text{else}
\end{cases}
\end{align}
and $T_k (x)$ is the $k^{th}$ degree Chebyshev polynomial of the first kind  \cite{subramanian2019Implementing}. 
This expansion is useful because a ``quantum walk" can exactly produce the effect of Chebyshev polynomials in $\hat{\mathcal{H}}/d$, 
where $d$ is the sparsity of the matrix. A quantum walk is a process in which the Hilbert space is enlarged, and then a walk operator 
that works in the enlarged Hilbert space is repeatedly implemented. For details on quantum walks and their implementations, 
see e.g. Refs.~\citenum{venegas2012quantum,kempe2003quantum}. 
With amplitude amplification to obtain a constant success probability of preparing a quantum state and a flag qubit to indicate 
if the preparation was successful, the query complexity of this method is $\mathcal{O}\left( n d^{n-1} \lambda_0^{-n}\right)$ 
 \cite{subramanian2019Implementing}.
Using this method without amplitude amplification, the query complexity of this implementation is $\mathcal{O}\left( n \right)$. 
Here, $ \lambda_0$ denotes the smallest (magnitude) eigenvalue of $H$, with the assumption that $\hat{\mathcal{H}}$ does not 
have 0 as an eigenvalue.
However, this method has its own drawbacks. The implementation of the quantum walk can be expensive, requiring a number of 
ancillary qubits that grows linearly with the system size \cite{subramanian2019Implementing}. 
The Chebyshev polynomial expansion's potential application for computing the imaginary time evolution of an unnormalized 
quantum state, and in general $e^{-\beta \hat{\mathcal{H}}}$ for evaluating partition function has been discussed 
in Ref. \citenum{Patel2018}, where they use the recursion relation for Chebyshev polynomials.

\subsection{Linear combination of unitary time propagator}

Recently, Seki and Yunoki proposed a quantum power method to evaluate $\hat{\mathcal{H}}^n|\phi\rangle$  \cite{seki2021quantum}, where the Hamiltonian power $\hat{\mathcal{H}}^n$ is first expressed as the $n$-th time derivative of $\hat{\mathcal{U}}(t)= e^{i\hat{\mathcal{H}}t}$ at $t=0$, then the time derivative is approximated through the central finite difference (CFD) which can be alternatively expressed as a linear combination of $\hat{\mathcal{U}}(\Delta_t)$'s at different time variables $\Delta_t$'s in close proximity to $t=0$, i.e.
\begin{align}
\hat{\mathcal{H}}^n &= i^n \frac{\partial^n \hat{\mathcal{U}}(t)}{\partial t^n} \bigg|_{t=0} 
\approx i^n \left[ \hat{\mathcal{U}}(\frac{\Delta_t}{2}) - \hat{\mathcal{U}}(-\frac{\Delta_t}{2}) \right]^n/\Delta_t^n \notag \\
&= \sum_{k=0}^n \left[\frac{i^n}{\Delta_t^n} (-1)^k  \Big(\substack{n\\k}\Big) \right] \left[ \hat{\mathcal{U}}(\frac{\Delta_t}{2}) \right]^{n-2k}. \label{time_der}
\end{align}
To implement the approach to quantum circuit, each $\hat{\mathcal{U}}(\frac{\Delta_t}{2})$ is further decomposed 
using symmetric Suzuki-Trotter decomposition (SSTD) \cite{trotter1959product,suzuki1990fractal}. 
There are two sources of error in this approach, the CFD error and SSTD error. The systematic CFD error scales 
quadratically with $\Delta_t$, i.e. $\mathcal{O}(\Delta_t^2)$, while the $2m$-th order SSTD brings the 
systematic error of $\mathcal{O}(\Delta_t^{2m})$. However, there are several advantages in the proposed approach. 
First, both aforementioned errors can be systematically suppressed through Richardson 
extrapolation \cite{richardson1927deferred,temme2017error} at the cost of involving more terms in the 
linear combination in Eq. (\ref{time_der}), which on the other hand offers the space for using low order SSTD. 
Second, the number of gates required for approximate $\hat{\mathcal{H}}^n$ for an $N_q$-qubit 
Hamiltonian $\hat{\mathcal{H}}$ scales as $\mathcal{O}(nN_q)$. In principle, the quantum power method can be 
combined with many classical approaches, which will be elaborated in Section \ref{classical} for targeting the 
ground/excited states and properties of a quantum many-body system. As shown in Ref. \citenum{seki2021quantum}, 
numerical noiseless simulations employing the quantum power method in Krylov-subspace diagonalization for 
targeting ground state of model systems demonstrate systematically improved accuracy over the conventional VQE. 
 
Inspired by the classical inverse power iteration approach for finding the dominant eigenstate of a given hermitian matrix with a more favorable logarithmic complexity in the iteration depth \cite{sachdeva2014faster}, Kyriienko recently proposed a quantum inverse iteration algorithm \cite{kyriienko2020quantum} for approximately computing the inverse of $n$-th power of the Hamiltonian, $\hat{\mathcal{H}}^{-n}$. The approach extends the discretized Fourier approximation of the Hamiltonian inverse \cite{childs2017quantum} to the $n$-th power,
\begin{align}
\hat{\mathcal{H}}^{-n} \approx \frac{iN_n}{\sqrt{2\pi}} \sum_{j=0}^{J} y_j^{n-1} \Delta_y \sum_{k=-K}^{K} z_k\Delta_z e^{-z_k^2/2} \hat{\mathcal{U}}_{j,k}, \label{Fourier}
\end{align}
where $N_n$ is a normalization factor, $y_j \equiv j\Delta_y$ and $z_k \equiv k \Delta_z$ with the integers $j\in [0,J]$ 
and $k\in [-K,K]$ and intervals $\Delta_{y,z}\in\mathbb{R}$ defining a discretization grid space, and unitary 
$\hat{\mathcal{U}}_{j,k}$ is given by $e^{-iy_jz_k\hat{\mathcal{H}}}$. Then $\langle \phi |\hat{\mathcal{H}}^{-n}|\phi\rangle$ 
is evaluated through the SWAP test \cite{ekert2002direct,Higgott2019variationalquantum}, overlap measurement \cite{mitarai2019methodology}, 
or Bell-type-like measurement \cite{kyriienko2020quantum}. However, this method requires a number of calls to a 
time-evolution oracle that is dependent on the condition number of the system, as well as a requirement that the 
Hamiltonian has been shifted such that all of the eigenvalues are positive. This latter requirement requires the 
user to have some knowledge of the ground state energy of the system \textit{a priori}. A suitable guess can be found, 
for example, using the minimum label finding algorithm of Ref.~\citenum{Ge2019}.

Alternatively, if considering a fault-tolerant implementation with favorable resource scaling $\hat{\mathcal{H}}^{n}|\phi\rangle$ or $\hat{\mathcal{H}}^{-n}|\phi\rangle$ can be done through amplitude amplification approach \cite{brassard2002quantum},  Hamiltonian simulation \cite{berry2015simulating}, qubitization \cite{low2017optimal}, or the direct block-encoding \cite{gilyen2019quantum} methods at the cost of introducing deeper circuit and implementing controlled-$\hat{\mathcal{U}}$ operations which, however, are still challenging to be implemented in the NISQ devices. Some of these methods have deep connection to linear combination of unitaries (LCU) technique \cite{Childs2012}. The basic idea of LCU can be briefly demonstrated from the following toy example. Suppose we want to implement  $\hat{\mathcal{U}}_0+\hat{\mathcal{U}}_1$ on $|\phi\rangle$. We can start by introducing an ancilla qubit and preparing  $|+\rangle|\phi\rangle$ with $|+\rangle \equiv \frac{1}{\sqrt{2}}(|0\rangle + |1\rangle)$. Then by performing the controlled unitary $|0\rangle\langle0|\otimes \hat{\mathcal{U}}_0+|1\rangle\langle 1|\hat{\mathcal{U}}_1$ on $|+\rangle|\phi\rangle$ we are able to obtain the state $\frac{1}{\sqrt{2}}(|0\rangle\hat{\mathcal{U}}_0|\phi\rangle+|1\rangle\hat{\mathcal{U}}_1|\phi\rangle)$. Now, if we measure the ancilla qubit in the $\{|+\rangle,|-\rangle\}$ basis and obtain the outcome associated with $|+\rangle$, then we obtain a state proportional to $(\hat{\mathcal{U}}_0+\hat{\mathcal{U}}_1)|\phi\rangle$.

\section{Classical computation of target states from quantum computed moments}\label{classical}

Now supposing we already have necessary Hamiltonian moments computed from quantum simulation, how can we proceed to accurately evaluate the energy and/or properties of the Hamiltonian? In this section, we will try to give our answers from some classical approaches.

\subsection{Lanczos approach}

One of the most common approaches that can be exploited is the Lanczos approach \cite{lanczos1950iteration}. In the Lanczos approach, an orthonormal set of quantum states $\{|\phi_i\rangle, i=1,2,\cdots\}$ is generated through a three-term recursion
\begin{align}
\hat{\mathcal{H}}|\phi_{i+1}\rangle = \beta_{i}|\phi_{i}\rangle+\alpha_{i+1}|\phi_{i+1}\rangle+\beta_{i+1}|\phi_{i+2}\rangle,
\end{align}
where $\alpha_i=\langle\phi_i|\hat{\mathcal{H}}|\phi_i\rangle$, $\beta_i=\langle\phi_i|\hat{\mathcal{H}}|\phi_{i+1}\rangle$. The three-term recursion essentially transforms the Hamiltonian $\hat{\mathcal{H}}$ into a tridiagonal form 
\begin{align}
\hat{\mathcal{T}}_i = \left[
\begin{array}{cccc}
\alpha_1 & \beta_1    &         &         \\
\beta_1 & \alpha_2    & \ddots  &         \\
    & \ddots & \ddots  & \beta_{i-1} \\
    &        & \beta_{i-1} & \alpha_i
\end{array}
\right] \label{T}
\end{align}
where the matrix elements $\alpha_i$ and $\beta_i$ can essentially be represented recursively in terms of Hamiltonian moments. Explicitly, it can be shown
\begin{align}
\alpha_1 &= \langle\hat{\mathcal{H}}\rangle, \notag \\
\beta_1 &= \sqrt{\langle\hat{\mathcal{H}}^2\rangle-\langle\hat{\mathcal{H}}\rangle^2}, \notag \\
\alpha_2 &= \frac{\langle\hat{\mathcal{H}}^3\rangle-2\langle\hat{\mathcal{H}}^2\rangle\langle\hat{\mathcal{H}}\rangle+\langle\hat{\mathcal{H}}\rangle^3}{\langle\hat{\mathcal{H}}^2\rangle-\langle\hat{\mathcal{H}}\rangle^2}, \notag \\
\beta_2 &= \frac{\langle\hat{\mathcal{H}}^4\rangle-(\alpha_1+\alpha_2)^2\beta_1^2-(\alpha_1^2+\beta_1^2)^2}{\beta_1^2}, \label{explicit_Lanczos} \\
\vdots \notag
\end{align}
Now the strict upper bound of the ground state energy of the Hamiltonian can be obtained by directly diagonalizing $\hat{\mathcal{T}}_i$. It has been shown that the classical Lanczos scheme can directly be applied even employing the noisy $\langle \hat{\mathcal{H}}^n \rangle$ \cite{Suchsland2021algorithmicerror}, which effectively provides a way to improve the energy estimate from real quantum hardware without being subject to specific hardware and an explicit description of the underlying noise. Preliminary tests on H$_2$, H$_3$, and four-site tetrahedral Heisenberg model demonstrate that combining the Lanczos scheme with VQE algorithm helps correct the noise and enhance the quality of the ansatz.

Alternatively, the matrix elements $\alpha_i$ and $\beta_i$ can be expressed in terms of the so-called connected moments (or cumulant) \cite{horn1984t}. An $n$-th order connected moments is defined as
\begin{align}
    \langle \hat{\mathcal{H}}^{n} \rangle_c \equiv \langle \phi | \hat{\mathcal{H}}^n | \phi \rangle - \sum_{m=0}^{n-1} \left(
    \begin{array}{c} n-1\\m-1\end{array}\right)
    \langle \hat{\mathcal{H}}^{m} \rangle_c \langle \phi | \hat{\mathcal{H}}^{n-m} | \phi \rangle, \label{connected_moment}
\end{align}
from which (\ref{explicit_Lanczos}) can be generalized as $z$-expansions of $\alpha$ and $\beta$ \cite{hollenberg1993plaquette,witte1994plaquette,hollenberg1996analytic}
\begin{align}
\alpha(z) &= \langle \hat{\mathcal{H}} \rangle_c + z \left( \frac{\langle \hat{\mathcal{H}}^{3} \rangle_c}{\langle \hat{\mathcal{H}}^{2} \rangle_c} \right) + 
\mathcal{O}(z^2), \notag \\
\beta^2(z) &= z\langle \hat{\mathcal{H}}^{2} \rangle_c + z^2\left(\frac{\langle \hat{\mathcal{H}}^{2} \rangle_c\langle \hat{\mathcal{H}}^{4} \rangle_c-\langle \hat{\mathcal{H}}^{3} \rangle_c^2}{2\langle \hat{\mathcal{H}}^{2} \rangle_c^2}\right) 
+\mathcal{O}(z^3), \label{general_lanczos}
\end{align}
where $z\equiv i/V$ with $i$ the recursion index and $V$ the volume of the system. From the bound analysis of orthogonal polynomials \cite{vandoorn1987representations,ismail1992bound}, it has been shown that the true ground state energy of the Hamiltonian can be approximated through the greatest lower bound (i.e. infimum) of the expansion (\ref{general_lanczos}) \cite{hollenberg1996analytic},
\begin{align}
E_0 \approx E_0^{\rm inf} = \inf_{z>0} [\alpha(z) - 2\beta(z)]. \label{infimum}
\end{align}
For example, the first order in $z$ gives \cite{hollenberg1994general}
\begin{align}
&E_0^{\rm inf} = \langle \hat{\mathcal{H}} \rangle_c - \frac{\langle \hat{\mathcal{H}}^2 \rangle_c^2}{\langle \hat{\mathcal{H}}^3 \rangle_c^2 - \langle \hat{\mathcal{H}}^2 \rangle_c\langle \hat{\mathcal{H}}^4 \rangle_c} \times \notag \\
&~~~~~~~~~~~~~~~~\Bigg( \sqrt{3\langle \hat{\mathcal{H}}^3 \rangle_c^2 - 2\langle \hat{\mathcal{H}}^2 \rangle_c\langle \hat{\mathcal{H}}^4 \rangle_c} - \langle \hat{\mathcal{H}}^3 \rangle_c\Bigg). \label{1st_inf}
\end{align}
Approach (\ref{1st_inf}) has been employed to demonstrate the energy estimate of 2D quantum magnetism model on 
lattices up to 25 qubits on IBM-Q quantum hardware \cite{Vallury2020quantumcomputed}, where the results show a 
consistent improvement of the estimated energies in comparison to the VQE results for the same ansatz.
Another application of the Lanczos method is the continued fractions expansion of the Green's function. 
The Green's function can be calculated in the Lehmann representation using the $\alpha$ and $\beta$ parameters 
from the Lanczos method. For full details, see e.g. Chapter 8 of Ref.~\citenum{koch2011}.

It is worth mentioning that standard Lanczos is a variational approach while the infimum approach (\ref{infimum}) does not offer a strict upper bound of the true ground state energy, which, as exemplified by (\ref{1st_inf}), is essentially an alternative size-extensive connected moment expansion of the ground state energy. Other polynomial expansion approaches, in particular connected moment expansions, will be discussed in details in Section \ref{poly_exp}.


\subsection{Real time evolution} \label{diagonalization}

Alternatively, one can resort to time evolution approaches to target the energy of the Hamiltonian. 
Here the time evolution approaches include both real time evolution (RTE) approach and its imaginary analogue. 
In RTE, the dynamics of a quantum state $|\phi_t\rangle$ is governed by the time-dependent Schr\"{o}dinger equation 
\begin{align}
    i \frac{\partial |\phi_t\rangle}{\partial t} = \hat{\mathcal{H}} |\phi_t\rangle. \label{TDSE}
\end{align}
In the Schr\"{o}dinger picture the solution of Eq. (\ref{TDSE}), $|\phi_t\rangle$, can be expressed by acting a time-evolution operator $\hat{U}(t) = e^{-i\hat{H}t}$ (or its truncated Taylor expansion in practice) on an initial state $|\phi_0\rangle$,
\begin{align}
    |\phi_t\rangle &= \hat{U}(t) |\phi_0\rangle \notag \\
    &= e^{-i \hat{\mathcal{H}} t} |\phi_0\rangle = \sum_{n=0}^{+\infty} \frac{(-it)^n}{n!} \hat{\mathcal{H}}^n |\phi_0\rangle. \label{RT_state}
\end{align}
The introduction of $\hat{\mathcal{H}}^n$ in (\ref{RT_state}) effectively builds a non-orthogonal Krylov subspace. Now consider an order-($r+1$) Krylov subspace
$\mathcal{K}_{r+1}=\{\hat{\mathcal{H}}^n|\phi_0\rangle,n=0,1,\cdots,r\}$,
the ground state can then be approximated as a linear combination of the Krylov basis
\begin{align}
	|\phi_t\rangle &= \sum_{n=0}^r c_n \hat{\mathcal{H}}^n|\phi_0\rangle \label{ansatz}
\end{align}
under the constraint $\langle \phi_t | \phi_t\rangle = 1$ with $c_n$'s being the expansion coefficients to be determined.
Plugging the expanded form (\ref{ansatz}) to Eq. (\ref{TDSE}) and projecting the approximate evolution on to the subspace $\mathcal{K}_{r+1}$, we end up with a set of time-dependent coupled equations in terms of the first $2r+1$ Hamiltonian moments, which in matrix form can be expressed as
\begin{align}
\dot{\mathbf{c}} = -i \mathbf{L}^{-1} \mathbf{R} \mathbf{c} \label{TDCE}
\end{align}
where the vector $\mathbf{c} = (c_0, c_1, \cdots, c_n)^T$ and $\dot{\mathbf{c}}$ is its time-derivative. Hamiltonian moments matrices $\mathbf{L}$ and $\mathbf{R}$ are defined as follows,
\begin{align}
\mathbf{L} &=
\left(\begin{array}{cccc}
1 & \langle \hat{\mathcal{H}} \rangle & \cdots & \langle \hat{\mathcal{H}}^r \rangle \\
\langle \hat{\mathcal{H}} \rangle & \langle \hat{\mathcal{H}}^2 \rangle & \cdots & \langle \hat{\mathcal{H}}^{r+1} \rangle \\
\vdots & \vdots &\ddots &\vdots \\
\langle \hat{\mathcal{H}}^{r} \rangle & \langle \hat{\mathcal{H}}^{r+1} \rangle & \cdots & \langle \hat{\mathcal{H}}^{2r} \rangle
\end{array}\right) \notag \\
\mathbf{R} &=
\left(\begin{array}{cccc}
\langle \hat{\mathcal{H}} \rangle & \langle \hat{\mathcal{H}}^2 \rangle & \cdots & \langle \hat{\mathcal{H}}^{r+1} \rangle \\
\langle \hat{\mathcal{H}}^2 \rangle & \langle \hat{\mathcal{H}}^3 \rangle & \cdots & \langle \hat{\mathcal{H}}^{r+2} \rangle \\
\vdots & \vdots &\ddots &\vdots \\
\langle \hat{\mathcal{H}}^{r+1} \rangle & \langle \hat{\mathcal{H}}^{r+2} \rangle & \cdots & \langle \hat{\mathcal{H}}^{2r+1} \rangle
\end{array}\right). \notag
\end{align}
%
Eq. (\ref{TDCE}) is a first order, linear ordinary differential equation (ODE), whose solution depends on the eigenvalues of the matrix $\mathbf{L}^{-1} \mathbf{R}$ that essentially approximate some of the eigenvalues of $\hat{\mathcal{H}}$. In a recent study \cite{guzman2021predicting},  this approach is demonstrate to be able to to target both the ground and excited states for pairing Hamiltonian. Alternatively, as shown in Ref.  \citenum{seki2021quantum}, employing Rayleigh-Ritz technique, $\mathbf{L}$ and $\mathbf{R}$ constitute general eigenvalue problem
\begin{align}
\mathbf{R} \mathbf{v} = \epsilon \mathbf{L} \mathbf{v} \label{eigen}
\end{align}
with the eigenpair ($\epsilon, \mathbf{v}$) approximate the energy and state vector of the true ground state. To solve Eq. (\ref{eigen}) for ($\epsilon, \mathbf{v}$), employ canonical orthogonalization to do the following transformation
\begin{align}
\mathbf{U}^\dagger \mathbf{R} \mathbf{U} (\mathbf{U}^{-1} \mathbf{v}) = \epsilon \mathbf{U}^\dagger \mathbf{R} \mathbf{v} = \epsilon (\mathbf{U}^{-1} \mathbf{v}),
\end{align}
where $\mathbf{U} \mathbf{U}^\dagger = \mathbf{L}^{-1}$. Since $\mathbf{L}$ is a hermitian that can be diagonalized by a 
unitary matrix $\mathbf{V}$ with eigenvalues constituting a diagonal matrix $\mathbf{s}$, i.e. 
$\mathbf{L} = \mathbf{V} \mathbf{s} \mathbf{V}^\dagger$, $\mathbf{U}$ can be constructed through 
$\mathbf{U} = \mathbf{V} \mathbf{s}^{-1/2}$. Similar subspace diagonalization schemes have been 
adopted in many other hybrid quantum-classical algorithms such as quantum subspace expansion 
\cite{mcclean2017hybrid,colless2018computation}, and some variants of VQE \cite{parrish2019quantum2,nakanishi2019subspace,huggins2020non}.


\subsection{Imaginary time evolution} \label{poly_exp}

The imaginary time evolution (ITE) approach has a long history of being a robust computational approach to solve the ground state of a many-body quantum system. 
To see how it works, let's first assume the quantum state at time $t$, $|\phi_t\rangle$, is now expanded in the eigen-space of $\hat{\mathcal{H}}$, $\{|\psi_n\rangle,n=0,1,\cdots\}$, with $E_n$'s being the corresponding eigenvalues, then as we show in the previous section $|\phi_t\rangle$ can be expressed as
\begin{align}
    |\phi_t\rangle = \sum_n c_n e^{-i E_n t} |\psi_n\rangle \label{RTE}
\end{align}
with $c_n$'s being the expansion coefficients at the initial time. Now, if we consider a replacement $\tau = it$ ($\tau$ is often called imaginary time) and confine $\tau\geq 0$, Eq. (\ref{RTE}) becomes
\begin{align}
    |\phi_{\tau}\rangle = \sum_n c_n e^{- E_n (\tau-\tau_0)} |\psi_n\rangle. \label{ITE}
\end{align}
Comparing Eqs. (\ref{RTE}) and (\ref{ITE}), we see that the wavefunction is driven from an ``oscillating" superposition of the Hamiltonian eigenstates to an ``exponential decaying" superposition of the eigenstates with the decay rate proportional to $E_n$. More importantly, in the limit of large $\tau$
\begin{align}
    |\phi_{\tau \gg 1}\rangle \approx c_0 e^{-E_0 \tau} |\psi_0\rangle,
\end{align}
the ground state $\psi_0(x)$ is ``screened out" because its exponential decay rate is the smallest in the eigen spectrum of $\hat{\mathcal{H}}$. Briefly speaking, given a trial wavefunction $|\phi\rangle$ that has non-zero overlap with the true ground state wavefunction $\psi_0$, $\langle \phi | \psi_0 \rangle \neq 0$, the ITE approach, in comparison to the RTE of the trial state, guarantees a monotonically decreasing energy functional in $\tau$ that converges to the ground state energy at $\tau \rightarrow \infty$, i.e.
\begin{align}
E_0 = \lim_{\tau\rightarrow +\infty} E(\tau) ~~\text{with}~~ E(\tau) = \langle \phi_\tau | \hat{\mathcal{H}} | \phi_\tau \rangle. \label{E_ITE}
\end{align}
Here the ITE of the trial wavefunction, $|\phi_t\rangle$, is defined as
\begin{align}
    |\phi_t\rangle = \frac {e^{-\tau \hat{\mathcal{H}}/2} | \phi \rangle}{\left(\langle \phi | e^{-\tau \hat{\mathcal{H}}} | \phi \rangle\right)^{1/2}}. \label{IT_state}
\end{align}
and the first derivative of $E(\tau)$ with respect to $\tau$ is non-positive
\begin{align}
    \frac{\rm{d}E(\tau)}{\rm{d}\tau} = -\big( 
    \langle \phi_t | \hat{\mathcal{H}}^2 | \phi_t \rangle -
    \langle \phi_t | \hat{\mathcal{H}} | \phi_t \rangle^2
    \big) \leq 0, \label{derivative}
\end{align}
where the equality sign holds if and only if $|\phi_t\rangle$ is an exact eigenstate of $\hat{\mathcal{H}}$ (e.g. $|\psi_0\rangle$).

The development and application of ITE approach targeting the ground state wave function and energy dates back to 1970s, when the similar random-walk imaginary-time technique were developed for diffusion Monte-Carlo methods \cite{davies1980application, anderson1975random, anderson1979quantum, anderson1980quantum}.
Later on, in the pursuit of non-perturbative analytical tool for a wide variety of Hamiltonian systems that can be systematically improved, various polynomial expansions of Eq. (\ref{E_ITE}) have been studied extensively. For example, a straightforward way is to utilize a finite number of Hamiltonian moments, $\hat{\mathcal{H}}^n$, to constitute a truncated Taylor expansion of $E(\tau)$ which we will have a detailed examination in Section \ref{simulation}. \\

\noindent \textbf{t-expansion} Beyond the conventional Taylor expansion, to reproduce asymptotic behavior of $E(\tau)$ over a longer imaginary time, Horn and Weinstein introduced a power series expansion in imaginary time $\tau$ for $E(\tau)$ in the early 1980s \cite{horn1984t},
\begin{align}
   E(\tau) =  \sum_{n=0}^{\infty} \frac{(-\tau)^n}{n!} \langle \hat{\mathcal{H}}^{n+1} \rangle_c, \label{t_exp}  
\end{align}
where $\langle \hat{\mathcal{H}}^{n+1} \rangle_c$ is the connected moments defined in Eq. (\ref{connected_moment}). 
The $\tau\rightarrow \infty$ behavior of $E(\tau)$ was then reconstructed through Pad\'{e} approximants to Eq. (\ref{t_exp}). Furthermore, it was found that it might be more efficient and error-resilient to first construct ($L,L+M$)-Pad\'{e} approximants ($M\geq2$) to $ \frac{\rm{d}E(\tau)}{\rm{d}\tau}$ rather than to $E(\tau)$, and then integrate the Pad\'{e} approximant from 0 to the critical $\tau$-value (where Eq. (\ref{derivative}) becomes positive) to obtain a larger set of approximations to $E(\tau)$. Early applications of the Pad\'{e} approximants of Eq. (\ref{t_exp}) on Heisenberg and Ising models demonstrated remarkable improvement upon mean-field results. \\

\noindent \textbf{Connected moment expansion} Based on the Horn-Weinstein theorem, Cioslowski further derived a more practical re-summation technique, the so-called connected moment expansion (CMX), to Eq. (\ref{t_exp}) to address the algebraic $\tau$-independent form of $E(\tau)$ \cite{cioslowski1987connected},
\begin{widetext}
\begin{eqnarray}
    \lim_{\tau\rightarrow +\infty} E(\tau)= \lim_{n\rightarrow \infty} E_0^{\rm CMX(\it n)} = \lim_{n\rightarrow \infty} \left(
    I_1-\frac{S_{2,1}^2}{S_{3,1}}\left(
    1+\frac{S_{2,2}^2}{S_{2,1}^2S_{3,2}}\left( 
    1+\cdots\left(
    1+\frac{S_{2,n-1}^2}{S_{2,n-2}^2S_{3,n-1}}
    \right)\cdots \right) \right) 
    \right), \label{cmx_recursion}
\end{eqnarray}
\end{widetext}
where the $S_{n,i}$'s are defined from the recursion
\begin{align}
    S_{n,1}   &= \langle \hat{\mathcal{H}}^n \rangle_c, ~~n=2,3,\ldots, \notag \\
    S_{n,i+1} &= S_{n,i}S_{n+2,i}-S_{n+1,i}^2, ~~ i\geq 1 . \notag
\end{align}
It is worth mentioning that further analysis of the CMX recursion had suggested that Eq. (\ref{cmx_recursion}) might be recast in a compact matrix form \cite{knowles1987validity,stubbins1988methods}
\begin{align}
    \lim_{n\rightarrow \infty} E^{{\rm CMX}(n)} 
    &= \langle \hat{\mathcal{H}} \rangle_c -
    \lim_{n\rightarrow \infty} \left( \langle \hat{\mathcal{H}}^2 \rangle_c ~~ \cdots ~~ \langle \hat{\mathcal{H}}^n \rangle_c \right) \mathbf{X}
\end{align} 
where vector $\mathbf{X}$ is the solution of the following linear system
\begin{align}
    \left(\begin{array}{ccc}
     \langle \hat{\mathcal{H}}^3 \rangle_c    & \cdots & \langle \hat{\mathcal{H}}^{n+1} \rangle_c\\
     \vdots & \ddots & \vdots \\
     \langle \hat{\mathcal{H}}^{n+1} \rangle_c    & \cdots & \langle \hat{\mathcal{H}}^{2n-1} \rangle_c
    \end{array}\right) \mathbf{X} = 
    \left(\begin{array}{c}
     \langle \hat{\mathcal{H}}^2 \rangle_c \\ \vdots \\ \langle \hat{\mathcal{H}}^n \rangle_c
    \end{array}\right). \label{cmx_mat}
\end{align}
The analytical properties of CMX and its comparison with other methods (e.g. Lanczos approach) have been extensively discussed in the literature \cite{knowles1987validity,prie1994relation,mancini1994analytic,ullah1995removal,mancini1995avoidance,fessatidis2006generalized,fessatidis2010analytic}. In particular, similar to perturbational theories, CMX is conceptually simple and size-extensive, and the accuracy can be easily tuned through the rank of the connected moments included in the approximation and/or the quality of the trial wave function. For the latter, there have been discussions on using multi-configurational \cite{cioslowski1987connected} or even correlated wavefunction (e.g. truncated CI or CC wave functions \cite{noga2002use}) in the CMX framework for accelerating the CMX convergence rate. Nevertheless, there were two major problems associated with the CMX calculations. First, the algebraic structure of the series expansion in the CMX could cause singularity \cite{mancini1994analytic}. Early efforts tried to address this issue with limited success by employing alternative moments expansion \cite{mancini1994analytic,ullah1995removal, mancini1995avoidance}, generalized moments expansion \cite{fessatidis2006generalized}, or generalized Pad\'{e} expansion \cite{knowles1987validity}. Second, CMX results are not variational. In comparison with the variational Lanczos methods \cite{mancini1991approximations,prie1994relation}, it was found that CMX might be considered as a limiting case of the strict Lanczos scheme, and their exact equivalence only holds in certain regions of parameter space. A systematic analysis of low order Lanczos and CMX ground state energy for model Hamiltonians such as harmonic oscilaltor, anharmonic oscillator, and Kondo model concluded that the accuracy of both approaches was dependent on the region of parameter space being studied \cite{prie1994relation}.\\

\noindent \textbf{Peeters-Devreese-Soldatov approach} Regarding the variational expansion, 
there had been another interesting yet less known moment approach proposed by Peeters and 
Devreese \cite{peeters1984upper}, and further analyzed by 
Soldatov \cite{soldatov1995generalized} in the 1990s, which we will refer to as the PDS approach. 
This approach originates in generalizing the Bogolubov inequality \cite{bogoliubov1947theory} 
and the Feynman inequality \cite{feynman1955slow} through the analysis of their 
Laplace transforms to obtain the upper bounds for the free energy. In particular, 
in the operator formalism, given a system Hamiltonian operator $\hat{\mathcal{H}}$ 
with ground state $E_0$, and a complex scalar $s\in \mathbb{C}$ with $\Re (s) > - E_0$, 
we can write the following inequality \cite{peeters1984upper,devreese1975self}
\begin{align}
    \langle e^{-\hat{\mathcal{H}}} \rangle \geq e^{-\langle \hat{\mathcal{H}} \rangle} \underset{\mathcal{L}^{-1}}{\stackrel{\mathcal{L}}{\Longleftrightarrow}} \left\langle \frac{1}{s+\hat{\mathcal{H}}} \right\rangle \geq \frac{1}{s+\langle \hat{\mathcal{H}} \rangle}.
\end{align}
with $\mathcal{L}$ and $\mathcal{L}^{-1}$ being the Laplace transform and inverse Laplace transform operators, and $\langle \cdot \rangle$ the expectation value with respect to a given trial wavefunction $|\phi\rangle$. The PDS formalism is based on expanding and analyzing $\left\langle\frac{1}{s+\hat{\mathcal{H}}}\right\rangle$ using  a simple identity (with introducing a parameter $a_1 \in \mathbb{R}$, and $a_1 \neq -s $)
\begin{align} 
\frac{1}{s+\hat{\mathcal{H}}}
&=\frac{1}{s+a_1}-\frac{\hat{\mathcal{H}}-a_1}{(s+a_1)^2} + \frac{1}{s+\hat{\mathcal{H}}} \cdot \frac{(\hat{\mathcal{H}}-a_1)^2}{(s+a_1)^2} . \;
\label{iden1}
\end{align}
Remarkably, the expectation value of the third term on the right hand side of identity (\ref{iden1}), defined as a residual term $R_1(s,a_1)$, is non-negative,
\begin{align}
    R_{1}(s,a_1) = \left\langle \frac{1}{s+\hat{\mathcal{H}}} \cdot \frac{(\hat{\mathcal{H}}-a_1)^2}{(s+a_1)^2} \right\rangle \geq 0, ~~(\Re(s) > -E_0). \label{R1_min}
\end{align}
By analyzing its first and second derivatives with respect to the induced parameter $a_1$, it is easy to show that when $a_1 = \langle \hat{\mathcal{H}} \rangle$, $R_1(s,a_1)$ reaches its local minima.

Now if we recursively expand $\frac{1}{s+\hat{\mathcal{H}}}$ in (\ref{iden1}) one more time, and introducing another parameter parameter $a_2 \in \mathbb{R}$, and $a_2 \neq -s $, we have
\begin{align} 
\frac{1}{s+\hat{\mathcal{H}}}
&=\frac{1}{s+a_1}-\frac{\hat{\mathcal{H}}-a_1}{(s+a_1)^2} + \left( \frac{1}{s+a_2}  
- \frac{\hat{\mathcal{H}}-a_2}{(s+a_2)^2} \right) \times \notag \\
&~~~~ \frac{(\hat{\mathcal{H}}-a_1)^2}{(s+a_1)^2}
+ R_2(s,a_1,a_2) \label{iden2}
\end{align}
with a refined residual term $R_2(s,a_1,a_2)$ also being non-negative for $\Re (s) > - E_0$
\begin{align}
R_2(s,a_1,a_2) &= \left\langle \frac{1}{s+\hat{\mathcal{H}}} \cdot \frac{(\hat{\mathcal{H}}-a_1)^2}{(s+a_1)^2} \cdot \frac{(\hat{\mathcal{H}}-a_2)^2}{(s+a_2)^2} \right\rangle \geq 0.
\end{align}
Examining the first and second derivatives of $R_2(s,a_1,a_2)$ with respect to $a_1$, $a_2$ shows when
\begin{align}
\left\{\begin{array}{l}
\langle (\hat{\mathcal{H}}-a_1) \cdot (\hat{\mathcal{H}}-a_1) (\hat{\mathcal{H}}-a_2)\rangle = 0 \\
\langle (\hat{\mathcal{H}}-a_2) \cdot (\hat{\mathcal{H}}-a_1) (\hat{\mathcal{H}}-a_2)\rangle = 0
\end{array}
\right. ,\label{lin_sys}
\end{align}
$R_2$ reaches its local minima. 

Note that there is an important feature about $a_i$ ($i=1,2$), a little transform of (\ref{lin_sys}) shows
\begin{align}
    a_i = \frac{\displaystyle\left\langle \hat{\mathcal{H}} \prod_{\substack{j=1 \\ j \ne i}}^2 (\hat{\mathcal{H}}-a_j)^2 \right\rangle}{\displaystyle\left\langle \prod_{\substack{j=1 \\ j \ne i}}^2 (\hat{\mathcal{H}}-a_j)^2 \right\rangle}, ~~ (i=1,2),
\end{align}
from which it is straightforward to show
\begin{align}
    \langle\hat{\mathcal{H}}\rangle \geq a_i \geq E_0, ~~\text{and}~~ s+a_i \geq 0, (i=1,2), \label{var1}
\end{align}
therefore all the terms in (\ref{iden2}) are non-negative, and 
\begin{align}
    R_1(s,a_1) \geq R_2(s,a_1,a_2). \label{var2}
\end{align}
(\ref{var1}) and (\ref{var2}) tell us that by keeping recursively refining $\frac{1}{s+\hat{\mathcal{H}}}$ using identity (\ref{iden1}), we are able to get tighter upper bounds to $E_0$. This then constitutes the basic idea of PDS approach. For example, if we recursively expand (\ref{iden1}) $K$ times ($K\in \mathbb{N}, K\geq2$) with each time introducing a new parameter $a_i$, $i=1,\ldots,K$), we will get a more refined non-negative residual term $R_K(s,a_1,\ldots,a_K)$
\begin{align} 
R_K(s,a_1,\ldots,a_K) &= \left\langle
\frac{1}{s+\hat{\mathcal{H}}} \prod_{i=1}^K \frac{(\hat{\mathcal{H}}-a_i)^2}{(s+a_i)^2}
\right\rangle, \label{iden3}
\end{align}
and
\begin{align} 
R_{K-1}(s,a_1,\ldots,a_{K-1}) \geq R_K(s,a_1,\ldots,a_K).
\end{align}
The $K$-th order PDS formalism, PDS($K$), is then associated with determining the introduced $K$ real parameters 
$(a_1^{(K)},\ldots,a_K^{(K)})$ that minimize the value of $R_K$ through
\begin{equation}
    \frac{\partial R_K(s,a_1^{(K)},\ldots,a_K^{(K)})}{\partial a_i^{(K)}}=0,
     ~~(i=1,\ldots,K),
\end{equation}
which can be alternatively translated to
\begin{align}
    &\left\langle \prod_{j=1}^K (\hat{\mathcal{H}}-a_j^{(K)}) \prod_{\substack{m=1 \\ m \ne i}}^K (\hat{\mathcal{H}}-a_m^{(K)}) \right\rangle = 0, ~~(i=1,\ldots,K).
\end{align}
with each $a_i^{(K)}$ ($i=1,\ldots,K$) being an strict upper bound to $E_0$, and can be expressed as
\begin{align}
    \langle \hat{\mathcal{H}}\rangle >  \cdots > a_i^{(K-1)} > a_i^{(K)} = \frac{\displaystyle\left\langle \hat{\mathcal{H}} \prod_{\substack{j=1 \\ j \ne i}}^K (\hat{\mathcal{H}}-a_j^{(K)})^2 \right\rangle}{\displaystyle\left\langle \prod_{\substack{j=1 \\ j \ne i}}^K (\hat{\mathcal{H}}-a_j^{(K)})^2 \right\rangle} \geq E_0. 
\end{align}

To numerically solve $a_i^{(K)}$'s in a PDS($K$) approach, take Eq. (\ref{lin_sys}) as an exmaple, we can first express Eq. (\ref{lin_sys}) in a matrix form 
\begin{align}
\mathbf{S} \cdot \mathbf{M} \cdot \mathbf{X}=\mathbf{0} \label{mat2}
\end{align} 
with
\begin{align} 
\mathbf{S} &= 
    \left( \begin{array}{cc}
        1 & -a_1 \\ 1 & -a_2
    \end{array} \right) ,~~
\mathbf{M} = 
    \left( \begin{array}{ccc}
        \langle\hat{\mathcal{H}}^3\rangle & \langle\hat{\mathcal{H}}^2\rangle & \langle\hat{\mathcal{H}}\rangle \\
        \langle\hat{\mathcal{H}}^2\rangle & \langle\hat{\mathcal{H}}\rangle & 1
    \end{array} \right), \notag \\
\mathbf{X} &= 
    \left( \begin{array}{c}
        1 \\ -a_1-a_2 \\ a_1a_2
    \end{array} \right) ,~~
\mathbf{0} = 
    \left( \begin{array}{c} 0 \\ 0 \end{array} \right).
\end{align}
Then, if we assume $a_1\neq a_2$, then $\det|\mathbf{S}|\neq 0$ and $\mathbf{S}^{-1}$ exists, therefore we are able to act $\mathbf{S}^{-1}$ on both sides of (\ref{mat2}) to get
\begin{align}
    \mathbf{M} \cdot \mathbf{X}=\mathbf{0} 
    \Leftrightarrow
    \mathbf{\tilde{M}} \cdot \mathbf{\tilde{X}}=-\mathbf{\tilde{Y}} \label{mat3}
\end{align}
with
\begin{align}
\mathbf{\tilde{M}} &=     
    \left( \begin{array}{cc}
        \langle\hat{\mathcal{H}}^2\rangle & \langle\hat{\mathcal{H}}\rangle \\
        \langle\hat{\mathcal{H}}\rangle & 1
    \end{array} \right),~~
\mathbf{\tilde{X}} = 
    \left( \begin{array}{c}
        -a_1-a_2 \\ a_1a_2
    \end{array} \right) ,\notag \\
\mathbf{\tilde{Y}} &= 
    \left( \begin{array}{c} 
        \langle\hat{\mathcal{H}}^3\rangle \\ \langle\hat{\mathcal{H}}^2\rangle 
    \end{array} \right).
\end{align}
It can be easily seen that given $\langle\hat{\mathcal{H}}^n\rangle$ ($n=1,2,3$), $\mathbf{\tilde{X}}$ (or equivalently $\mathbf{X}$) can be solved from (\ref{mat3}), and $a_i$ ($i=1,2$) are then the solution of the polynomial 
\begin{align}
    \text{Poly}(a) = \sum_{i=0}^2 X_i a^{2-i}. \label{poly}
\end{align}

Now generalizing (\ref{mat2})-(\ref{poly}) we get the working equation for the PDS($K$) approach, where one need to first solve (\ref{mat3}) for an auxiliary vector $\mathbf{\tilde{X}} = (X_1,\cdots,X_K)^T$ with matrix elements defined as
\begin{align}
\left\{
\begin{array}{cl}
M_{ij} &= \langle H^{2K-i-j} \rangle  \\
Y_i &= \langle H^{2K-i} \rangle
\end{array}
\right.,~~ (i,j = 1,\cdots,K),
\end{align}
and then solve the following polynomial
\begin{equation}
\text{Poly}_K(a) = a^{K} + \sum_{i=1}^K \tilde{X}_i a^{K-i}, \label{poly2}   
\end{equation}
for $(a_1^{(K)},\ldots,a_K^{(K)})$ that provide upper bounds to the exact ground and excited state energies 
of the Hamiltonian characterized by either discrete or continuous spectral resolutions, or both.

\subsection{Variational simulation of real and imaginary time dynamics}

In the RTE and ITE, the target quantum state is approximated through the evolutions of the trial state as shown in Eqs. (\ref{RT_state}) and (\ref{IT_state}). In practice, the trial state can usually be prepared by applying a sequence of parametrized gates to the initial state $|0\rangle$. For example, we can write
\begin{align}
|\phi\rangle = |\phi(\vec{\theta})\rangle = U_n(\theta_n)\cdots U_k(\theta_k)\cdots U_1(\theta) |0\rangle,
\end{align}
where $U(\vec{\theta}) = U_n(\theta_n)\cdots U_k(\theta_k)\cdots U_1(\theta)$ with $\vec{\theta} = (\theta_1,\cdots,\theta_n)$ and $U_k(\theta_k)$ the $k$-th one-/two-qubit gate parametrized by rotation $\theta_k$. In the conventional VQE, based on the measurement outcome of the trial state, classical optimization routines are then employed to find optimal $\vec{\theta}$ that minimizes the cost function defined as the measurement outcome of the quantum state $|\phi(\vec{\theta})\rangle$,
\begin{align}
E_{\min} = \min_{\vec{\theta}} E(\vec{\theta})= \min_{\vec{\theta}}\langle \phi(\vec{\theta})| \hat{\mathcal{H}} |\phi(\vec{\theta})\rangle,
\end{align}
which gives a lowest upper bound to the true ground state state energy of the many-body Hamiltonian $\hat{\mathcal{H}}$ in the parameter space. Similar variational procedure can also be applied following the RTE and ITE approaches by replacing the $|\phi\rangle$ in the VQE with the time-evolved state approximated as a linear combination of Krylov bases as shown in Eqs. (\ref{RT_state}) and (\ref{IT_state}).
There are at least two advantages of combining variational optimization with the time evolution of the quantum state. First of all, since the variational optimization also contributes to bring down the cost function, the combination would help reduce the number of Krylov bases employed to evolve the trial state, which in turn helps reduce the measurement requirement. Second, the Krylov bases employed to improve the fidelity of the trial state helps restructure the potential energy surface in the same parameter space, which in turn offers an effective approach to navigate the dynamics to be free from getting trapped in the local minima of the old potential energy surface.

These advantages have been demonstrated by Peng and Kowalski in a recent study \cite{Peng2021}, where Eq. (\ref{poly}) is effectively treated as an alternative energy functional for the variational quantum solver whose analytical energy derivative $\nabla E(\vec{\theta})$ can be exploted to drive the optimization
\begin{align}
\vec{\theta}_{k+1} = \vec{\theta}_{k} - \eta \mathcal{R}^{-1}(\vec{\theta})\nabla E(\vec{\theta}). \label{dynamics}
\end{align}
Here $\eta$ is the learning rate for updating $\vec{\theta}$ at step $k$, and $\mathcal{R}(\vec{\theta}_k)$ is the Riemannian metric matrix at $\vec{\theta}_k$ that is flexible to characterize the singular point of the parameter space and is essentially related to the indistinguishability of $E(\vec{\theta})$ \cite{yamamoto2019natural}. Note that similar formulation is quite often used in, for example, gradient descent approach \cite{lemarechal2012cauchy} with the metric being often replaced by the Hessian matrix. Different from the conventional gradient descent approach, Eq. (\ref{dynamics}) has its origin in the general nonlinear optimization framework featuring natural gradient for targeting machine learning problems \cite{amari1998NG}. Here, the natural gradient accounts for the geometric structure of the parameter space, and is expressed as the product of the Fisher information matrix and the gradient of the cost function often with the attempt of circumventing the plateaus in the parameter space \cite{mcardle2019variational, yamamoto2019natural,Stokes2020quantumnatural}. Analogously, in the quantum computing, the Fisher information matrix is replaced by the quantum Fubini-Study metric to describe the curvature of the ansatz class. 

There have been discussions about deriving and applying quantum Fubini-Study metric in the quantum simulations, and comparing its performance with classical Fisher metric in some recent reports \cite{mcardle2019variational, yamamoto2019natural,Stokes2020quantumnatural}. For example, following Ref.  \citenum{mcardle2019variational} one can show that the RTE or ITE approach can be translated to a variational approach for a given ansatz $|\phi_\tau\rangle = |\phi(\vec{\theta}(\tau))\rangle$, $\vec{\theta}(\tau)\in\vec{\mathbb{R}}$ for targeting the target state $|\phi\rangle$. Take ITE approach as an example, following McLachlan's variational principle \cite{mclachlan1964variational}
\begin{align}
\delta \|\hat{\mathcal{Q}}|\phi_{\tau}\rangle\|= 0, \label{mclachlan}
\end{align}
where
\begin{align}
    \hat{\mathcal{Q}}\equiv\frac{\partial}{\partial \tau} + \hat{\mathcal{H}} - E(\tau) 
\end{align}
and
\begin{widetext}
\begin{align}
    \|\hat{\mathcal{Q}}|\phi_{\tau} \rangle \|^2 &= \langle \phi_{\tau} | \hat{\mathcal{Q}}^\dagger \hat{\mathcal{Q}}|\phi_{\tau}  \rangle \notag \\
    &= \sum_{i,j} \frac{\partial \langle \phi_{\tau} |}{\partial \theta_i} \frac{\partial | \phi_{\tau} \rangle }{\partial \theta_j} \dot\theta_i \dot\theta_j + \sum_i \left(\frac{\partial \langle \phi_{\tau} |}{\partial \theta_i} (\hat{\mathcal{H}}-E(\tau))|\phi_{\tau} \rangle
    + \langle \phi_{\tau} | (\hat{\mathcal{H}} - E(\tau)) |\frac{\partial |\phi_{\tau}\rangle}{\partial\theta_i} \right) \dot\theta_i 
    + \langle \phi_{\tau} | (\hat{\mathcal{H}} - E(\tau))^2 | \phi_{\tau} \rangle, 
\end{align}
\end{widetext}
with $\dot\theta_i \equiv \frac{\partial \theta_i}{\partial \tau}$, we have
\begin{widetext}
\begin{align}
    \frac{\partial \|\hat{\mathcal{Q}}|\phi_{\tau} \rangle \|^2 }{\partial \dot\theta_i} 
    &= \sum_j \left( \frac{\partial \langle \phi_\tau |}{\partial \theta_i} \frac{\partial | \phi_\tau \rangle}{\partial \theta_j} + 
    \frac{\partial \langle \phi_\tau |}{\partial \theta_j} \frac{\partial | \phi_\tau \rangle}{\partial \theta_i} \right)\dot\theta_j  
    +\frac{\partial \langle \phi_\tau |}{\partial \theta_i} (H-E_{\tau})|\phi_\tau \rangle + \langle \phi_\tau | (H - E{\tau}) |\frac{\partial |\phi_\tau\rangle}{\partial \theta_i} = 0. \label{Mclachlan1}
\end{align}
\end{widetext}
Note that since $|\phi_\tau\rangle$ is normalized, $\langle \phi_\tau | \phi_\tau \rangle = 1$, we also have
\begin{align}
    0 = \frac{\partial \langle \phi_\tau | \phi_\tau \rangle}{\partial \theta_i} =\frac{\partial \langle \phi_\tau | }{\partial \theta_i}| \phi_\tau \rangle + \langle \phi_\tau | \frac{\partial | \phi_\tau \rangle }{\partial \theta_i},
\end{align}
and Eq. (\ref{Mclachlan1}) can be simplified as a linear system in matrix form
\begin{align}
\mathcal{R} \dot{\vec{\theta}} = -\nabla E(\vec{\theta}) \label{ite_dynamics}
\end{align}
which is essentially the dynamics (\ref{dynamics}) with the matrix elements $\mathcal{R}_{ij}$ given by
\begin{align}
    \mathcal{R}_{ij} = \Re \left(\frac{\partial \langle \phi_\tau |}{\partial \theta_i} \frac{\partial |  \phi_\tau \rangle}{\partial \theta_j}\right). \label{ite_riemanninan}
\end{align}
Same dynamics and Riemannian metric might be derived from other variational principles. As shown in Ref.  \citenum{Yuan2019theoryofvariational}, classically there are three variational principles, namely the Dirac and Frenkel variational principle \cite{dirac1930note,frenkel1934wave}, the McLachlan variational principle \cite{mclachlan1964variational}, and the time-dependent variational principle \cite{kramer1981,broeckhove1988tdvp}, that lead to the same evolution equation. However, when parameters are confined to be real, the Dirac and Frenkel and the McLachlan variational principles are equivalent, while the time-dependent variational principle cannot lead to a nontrivial evolution of the parameters.

It is also worth mentioning that a caveat in the above derivation is that we implicitly assume $|\phi_\tau\rangle = |\phi\rangle$. If there is time-dependent global phase difference between $|\phi_\tau\rangle$ and $|\phi\rangle$, their time-derivatives can be very different and the dynamics (\ref{ite_dynamics}) would be incorrect \cite{mcardle2019variational}. In essence, the $\mathcal{R}$ defined in Eq. (\ref{ite_riemanninan}) is not gauge invariant thus not qualified for measuring the quantum distance \cite{cheng2013quantum}. The problem can be fixed by either explicitly introducing a time-dependent phase gate to the trial state, or defining a gauge invariant quantum geometric tensor
\begin{align}
\mathcal{R}_{ij} = \Re \left(\frac{\partial \langle \phi_\tau |}{\partial \theta_i} \frac{\partial |  \phi_\tau \rangle}{\partial \theta_j}\right)
- \frac{\partial \langle \phi_\tau |}{\partial \theta_i} |  \phi_\tau \rangle \langle \phi_\tau | \frac{\partial |  \phi_\tau \rangle}{\partial \theta_j}
 \label{ngd_riemannian}
\end{align}
which is essentially associated with quantum natural gradient \cite{Stokes2020quantumnatural}.

\section{Quantum simulation of a model Hamiltonian} \label{simulation}

In the preceding sections we have given a brief review of hybrid quantum-classical algorithms featuring quantum computed Hamiltonian moments and their classical post-processing approaches that are in principle capable of providing systematically improvable estimates for the target energy and quantum state. In this section, we will go over a simple tutor to demonstrate how one can use the algorithms of these kinds in practice for solving some practical problems. 

In this tutorial, we will study a model Hamiltonian
\begin{align}
    \hat{\mathcal{H}} = J \sum_{\langle i, j \rangle} \left( \sigma_x^i \sigma_x^j + \sigma_y^i \sigma_y^j \right) + U \sum_{\langle i, j \rangle} \sigma_z^i \sigma_z^j + B \sum_i \sigma_z^i \label{heisenberg}
\end{align}
which, as suggested in Ref.  \citenum{Smith2019}, when combining with different choices of scalars $J$, $U$, and $B$, abstracts a large range of physical topics such as the integrability \cite{Essler2016quench}, quantum magnetism \cite{vasiliev2018milstones} and many-body localization \cite{abanin2017recent,nandkishore2015manybody}. In the following we will show how to employ a Hamiltonian moment based hybrid quantum-classical approach to compute the energy and magnetism of this model Hamiltonian (\ref{heisenberg}) governing four sites mapped to four qubits. 

\subsection{Computation preparation}

To launch the quantum computation, we first need to prepare our trial state. The following hardware efficient ansatz is used for the preparation of our trial state
\begin{align}
    |\phi_0 \rangle = C_z^{0,1}C_z^{1,2}C_z^{2,3}R_y^0(\theta_0)R_y^2(\theta_1)R_y^3(\pi)|1100\rangle
\end{align}
where $C_z^{p,q}$ is a controlled-Z operation with control qubit $p$ and target qubit $q$, $R_y^p(\theta)$ is a rotation of $\theta$ on qubit $p$ around $y$-axis, and two rotations $(\theta_0,\theta_1) = (-2.0,1.0)$. We employ the hybrid ITE approach for our demonstration where the Hamiltonian moments are evaluated from IBM-Q NISQ quantum hardware, and the energy and magnetism computations are performed using classical ITE approach on quantum computed Hamiltonian moments. Specifically, the ITE propagator is approximated using a $m$-order Taylor expansion, 
\begin{align}
e^{-\tau \hat{\mathcal{H}}/2} \simeq \sum_{n=0}^m \frac{(-\tau)^n}{n!}\hat{\mathcal{H}}^n, 
\end{align}
and therefore $\langle e^{-\tau \hat{\mathcal{H}}/2} \hat{\mathcal{H}} e^{-\tau \hat{\mathcal{H}}/2} \rangle$ and $\langle e^{-\tau \hat{\mathcal{H}}} \rangle$ can be approximated as linear combinations of $\langle \hat{\mathcal{H}}^n \rangle$. 

\subsection{Quantum measurement}

As mentioned in preceding sections, since $\hat{\mathcal{H}}^n$ can be generally expressed as a linear combination of Pauli strings, $P_i$'s, $\langle \hat{\mathcal{H}}^n \rangle$ is then evaluated based on the measurements of $\langle P_i \rangle$'s.
In quantum computation, there are generally two ways of measuring observables, direct and indirect measurement. In the former, the measured state collapses to the measurement basis that one can freely choose. In the latter, the measured state is not completely destroyed. One simplest and important example is the Hadamard test \cite{aharonov2009polynomial}. In the Hadamard test of $\langle \phi |U|\phi\rangle$ in the computational basis for a trial state $|\phi\rangle$ governed by a unitary $U$, one is able to reuse the state $\frac{(I\pm U)|\phi\rangle}{\sqrt{2}}$ after the measurement. This property can be exploited for designing more efficient and less resource-demanding version of some quantum algorithms (for example, iterative version of the quantum phase estimation, see Refs.  \citenum{knill2007optimal,dob2007arbitrary}). Here for the demonstration we first employ the Hadamard test to measure each $\langle P_i\rangle$ that contributes $\langle \hat{\mathcal{H}}^n \rangle$, and then to show how to group $P_i$'s to enable simultaneous measurements. 

\begin{figure}[hp]
\centering
\includegraphics[scale=0.6]{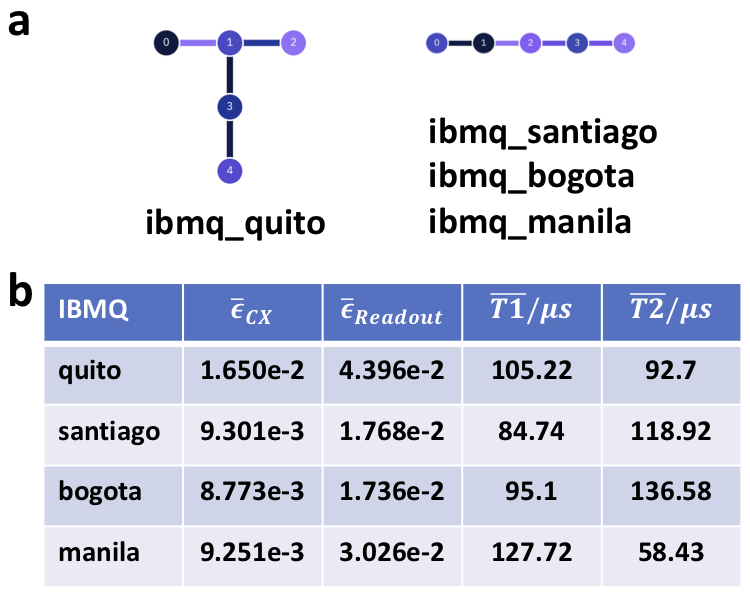}
\caption{(\textbf{a}) Topologies of four IBM-Q NISQ devices employed in this work. All the employed IBM-Q NISQ processors belong to Falcon family with basis gates being CX, ID, RZ, SX, and X.  The color of nodes implies frequency (GHz) of the qubit or how fast a 1-qubit gate can be executed. The color of the connection implies the gate time in nanoseconds for 2-qubit gate such as CX. (\textbf{b}) Average CX error ($\overline{\epsilon}_{CX}$), average readout error ($\overline{\epsilon}_{Readout}$), average T1/T2 time ($\overline{T1}/\overline{T2}$) of the employed IBM-Q devices.}  \label{ibmq}
\end{figure}

As discussed in Section \ref{measurement}, the Pauli cyclic rules (\ref{Pauli_reduction}) allow for some reduction of the number of Pauli strings that need to be measured. Here, for the four-site Heisenberg model, we found that 72 Pauli strings form a complete basis for reformulating arbitrary $\mathcal{H}^n$. This finding suggests that once $\langle P_i \rangle$'s are measured one can quickly (classically) compute approximate $\langle e^{-\tau \mathcal{H}} \rangle$ and $\langle e^{-\tau \mathcal{H}/2} \mathcal{H} e^{-\tau \mathcal{H}/2} \rangle$ using arbitrary order expansion for any $\tau$. Moreover, the relatively few number of required $P_i$'s allows the feasibility of measuring $\langle P_i\rangle$'s on physical quantum computers. Here, four IBM-Q's publicly available devices, {\it ibmq\_quito}, {\it ibmq\_santiago}, {\it ibmq\_bogota}, and {\it ibmq\_manila} are employed, whose topologies and noise metrics are shown in Fig. \ref{ibmq}. 

\subsection{Estimated ground state energy and magnetism}

We estimated the ground state energies of the four-site Heisenberg model Hamiltonian for varied $U$ and $J$ values with $B=1.0$ employing approximate ITE approach, where $e^{-\tau\hat{\mathcal{H}}}$ is approximated by a 15th order Taylor expansion. In all the calculations, we set $\tau=2.5$. A detailed discussion with respect to different truncation of the Taylor expansion and the choice of $\tau$ will be given in Section \ref{approximateITE}.

\begin{figure}[h]
\centering
\includegraphics[scale=0.25]{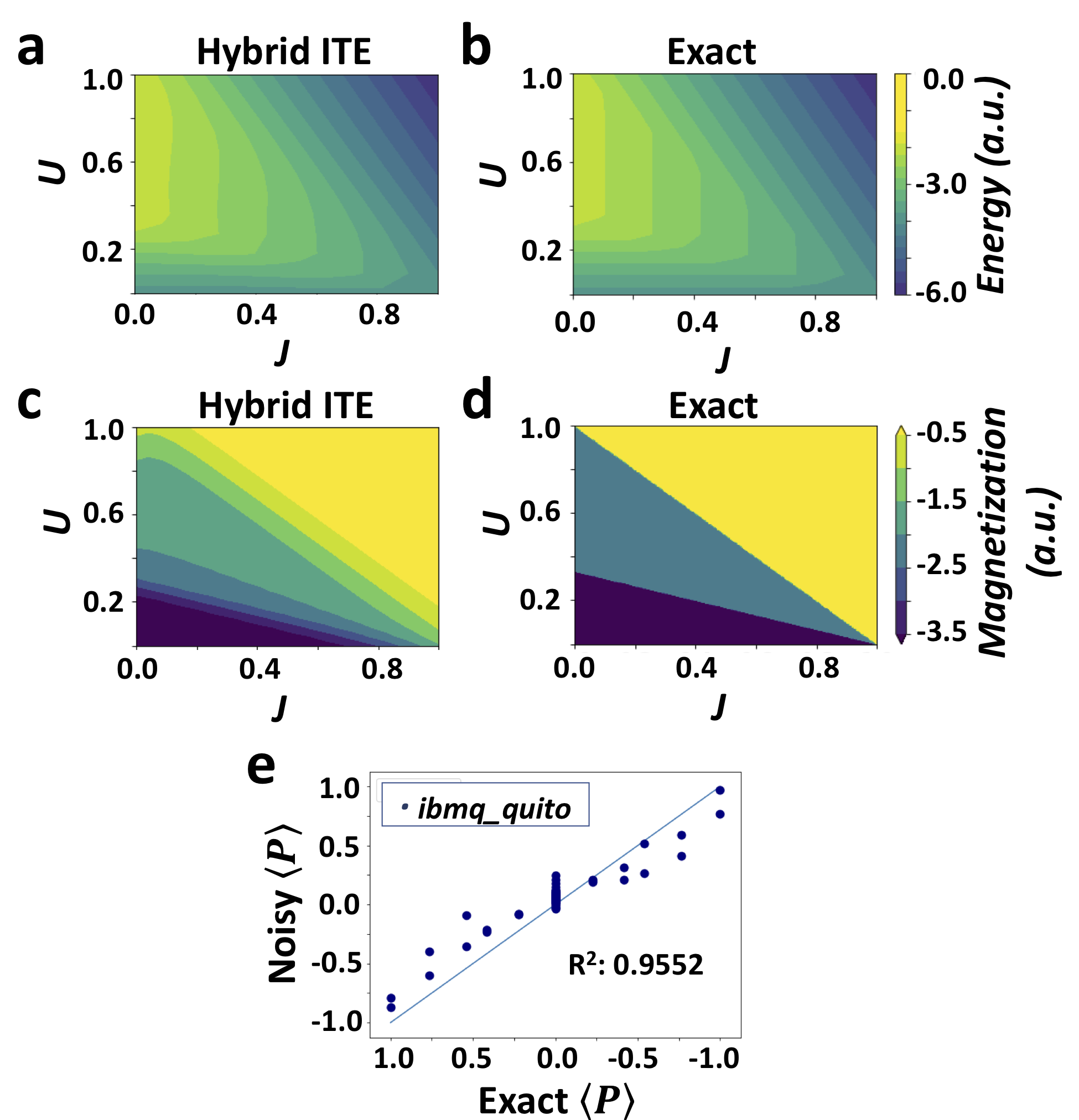}
\caption{(\textbf{a}) Ground state energies of the model Hamiltonian with varied $U$ and $J$ values estimated by the hybrid ITE approach, and (\textbf{b}) the corresponding exact ground state energies. (\textbf{c}) Magnetizations of the same model  estimated by the hybrid ITE approach, and (\textbf{d}) the corresponding exact magnetizations. In the hybrid ITE approach, a 15-th order Taylor expansion is employed to approximate $e^{-\tau \hat{\mathcal{H}}}$ with $\tau = 2.5$, and totally 72 $\langle P_i\rangle$'s have been measured on the quantum computer \textit{ibmq\_quito}. The measured $\langle P_i\rangle$'s are compared with their analytical values in (\textbf{e}). The Mean Squared Error between the estimated and exact ground state energies across over all the ($U$, $J$)'s is 0.0025 a.u.} \label{results}
\end{figure}

Fig. \ref{results}a exhibits the computed ground state energies of the model Hamiltonian with varied $\{U,J\}\in(0,1)$, which are in an excellent agreement with the exact solutions shown in Fig. \ref{results}b, and the average mean squared error between the two is only 0.0025 a.u. Based on the measured $\langle P_i\rangle$'s we are also able to evaluate the magnetization of the model Hamiltonian. The magnetization operator is given by $\hat{\mathcal{M}} = \sum_i \sigma_z^i$. The contour of the expectation values of magnetization with respect to the estimated ground states under different $U$ and $J$ values is shown in Fig. \ref{results}c, which are again in great agreement with the exact solutions shown in Fig. \ref{results}d. 

\subsection{Discussions}

\subsubsection{In comparison with VQE}

Remarkably, the ground state energy and magnetization estimations in the present study exhibit great improvement 
in comparison with the VQE solutions of the same model on the real hardware reported in previous work \cite{Kandala2017hardware}. 
It is also worth mentioning that (i) the ansatz we used here is much simpler and shallower, and thus less robust, 
than the one used in the previous study,  (ii) the measurement requirement for the model Hamiltonian is trivial 
in the present study, and does not heuristically depends on the quality of the ansatz and the number of iterations 
as it does in the conventional VQE practice, and (iii) as shown in Fig. \ref{results}e the deviation between 
measured and analytical $\langle P_i\rangle$'s could be as large as $\pm0.3\sim0.4$ a.u. given $\langle P_i\rangle\in(-1,1)$. 
Nevertheless, despite of less robust ansatz employed and the noisy $\langle P_i\rangle$'s, the ITE approach 
is able to give very accurate estimation based on trivial measurements, thus features a robust error mitigation ability 
from the algorithmic perspective. Similar feature has also been reported in the real quantum application of other Krylov 
subspace methods, for example the hybrid and quantum Lanczos approaches \cite{Suchsland2021algorithmicerror,motta2020determining}.

\subsubsection{Grouping and readout error mitigation} 

As discussed in the preceding section, we can further reduce the number of terms to be measured by grouping the Pauli strings that commute and performing simultaneous measurements \cite{Izmaylov2020}. Here, by applying QWC grouping, we found that the number of terms to be measured can be further reduced from 72 to 25 for the model Hamiltonian. 

Another advantage of performing simultaneous measurement is that the readout error of the simultaneous measurement could be largely mitigated through calibration. To see how it works, suppose we perform $n$-qubit simultaneous measurements for totally $2^n$ computational bases, the normalized resulting counts from the measurement of each basis, when collected, would constitute a matrix $\mathbf{J}$ of $2^n$ dimension. Under the ideal noise-free condition, we know that $\mathbf{J}$ is an identity matrix of the same dimension. In the real NISQ quantum computing, however, with the effect of noise $\mathbf{J}$ deviates from the identity matrix, and the extent of deviation implies how noisy the real measurement outcomes are. Based on this relation, if $\mathbf{J}$ is known, we can algebraically build a connection between any ideal counts vector $\mathbf{C}_{\rm ideal}$ and its noisy analogue $\mathbf{C}_{\rm noisy}$ from the real measurement through
\begin{align}
\mathbf{C}_{\rm ideal} = \mathbf{J}^{-1}\mathbf{C}_{\rm noisy}. \label{calibration}
\end{align}

\begin{figure}[h]
\centering
\includegraphics[scale=0.25]{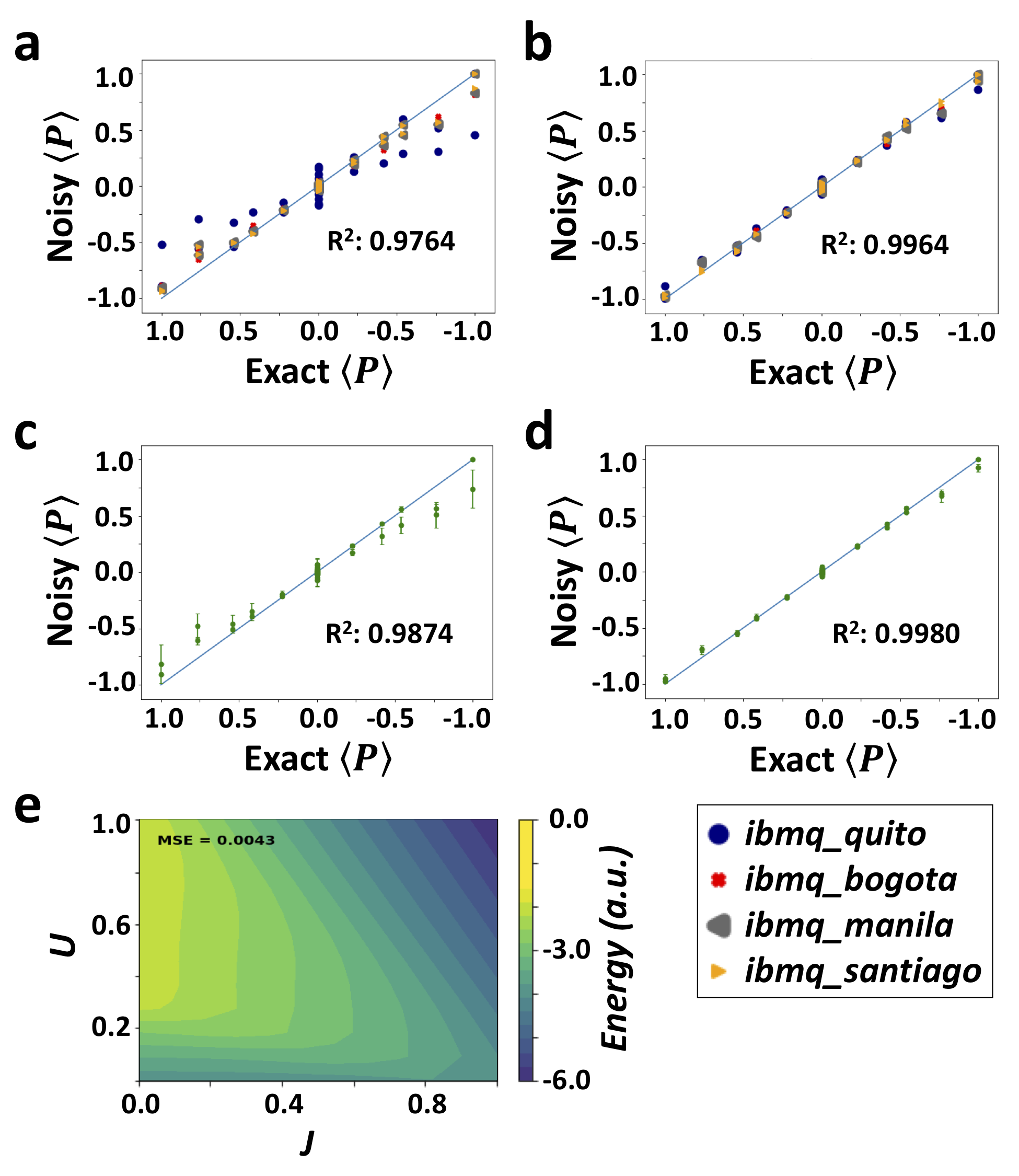}
\caption{(\textbf{a}) and (\textbf{b}) Expectation values of 25 QWC bases measured from four IBM-Q machines with five repeated executions on each machine without and with measurement readout calibration respectively. (\textbf{c}) and (\textbf{d}) Expectation values, as well as standard deviations, of 25 QWC bases averaged over all IBM-Q machines and runs without and with measurement readout calibration respectively. (\textbf{e}) The estimated ground state energies of the model Hamiltonian with varied $U$'s and $J$'s based on the calibrated simultaneous measurements of the 25 QWC bases.} \label{fig_calibration}
\end{figure}

Apparently, the construction of matrix $\mathbf{J}$ eats up the resource quickly due to the exponential growth of the dimension. Nevertheless, for small size problem like the four-site model Hamiltonian studied in the present work, $\mathbf{J}$ is a $16\times 16$ matrix, and a direction construction of $\mathbf{J}$ from simultaneously measuring 16 computation bases is still feasible. Here, after grouping the commuting Pauli strings into 25 QWC bases, we performed the simultaneous measurement in the computational basis for each QWC basis five times on four IBM-Q machines, and then mitigated the measurement outcomes employing (\ref{calibration}). The raw and calibrated measurement outcomes are exhibited in Fig. \ref{fig_calibration}a-d. As can be seen, after calibration the measurement outcomes are greatly improved showing great consistency with the analytical ones and significantly suppressing the deviation. Based on the measurement outcomes, the estimated ground state energies of the model Hamiltonian are in excellent agreement with the exact solutions (see Fig. \ref{fig_calibration}e), and the MSE between the estimated and exact energies over all the ($U,J$)'s is only 0.0045 a.u.

\subsubsection{Approximation in ITE} \label{approximateITE}

\begin{figure}[h]
\centering
\includegraphics[scale=0.25]{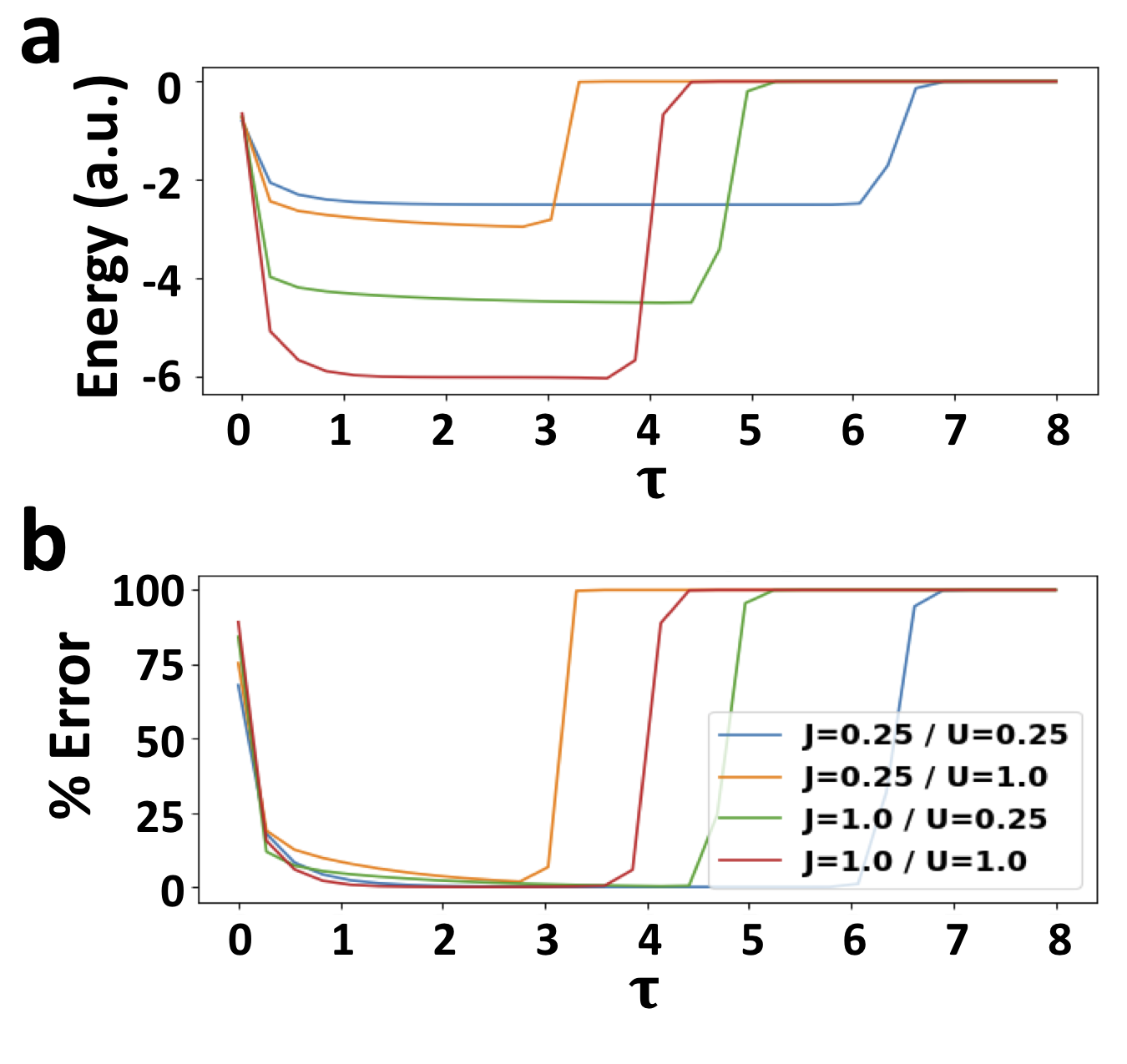}
\caption{(\textbf{a}) Energy evolutions and (\textbf{b}) corresponding error percentages of the four-site Heisenberg model Hamiltonian with different $U$'s and $J$'s as functions of imaginary time, $\tau$, estimated by the approximate ITE approach employing a 15-th order Taylor expansion.} \label{tau}
\end{figure}

In the case of the model Hamiltonian studied here, as stated at the beginning of this section we approximate the imaginary time operator $e^{-\tau \mathcal{H}}$ employing truncated Taylor expansion.
Generally speaking, for $x \in [0,b]$, the truncated expansion of an exponential function $e^{-x}$ is $\delta$-approximation when using $b+\log(\frac{1}{\delta})$ terms in the expansion \cite{sachdeva2014faster}. Regarding the ploynomial approximations to $e^{-\tau\hat{\mathcal{H}}}$, before figuring out the number of terms in the expansion, the variable $\tau$ needs to be determined first as it rescales the spectrum of $\hat{\mathcal{H}}$ and indirectly affects the level of accuracy of the employed expansion. To do so, as exemplified in Fig. \ref{tau}, we fix the number of terms in the expansion and evolve the energy as a function of $\tau$ to find its optimal value that gives the lowest energy. This is essentially a one dimensional optimization problem over $\tau$, and can be solved classically by Golden-Section Search  \cite{Kiefer1953}, or similar methods. In the present study, we found $\tau_{\rm opt}=2.5$ providing well converged ground state energies estimates for the given $U$ and $J$ ranges. It worth mentioning that to improve the estimate of $e^{-\tau\hat{\mathcal{H}}}$ using polynomial expansions in the larger $\tau$ regime, one can alternatively resort to other polynomial expansions, such as CMX as discussed in Section \ref{poly_exp}, or Pad\'{e} expansion as originally proposed in Ref.  \citenum{horn1984t} and recently applied in Ref.  \citenum{guzman2021predicting}.

\begin{figure}[h]
\centering
\includegraphics[scale=0.25]{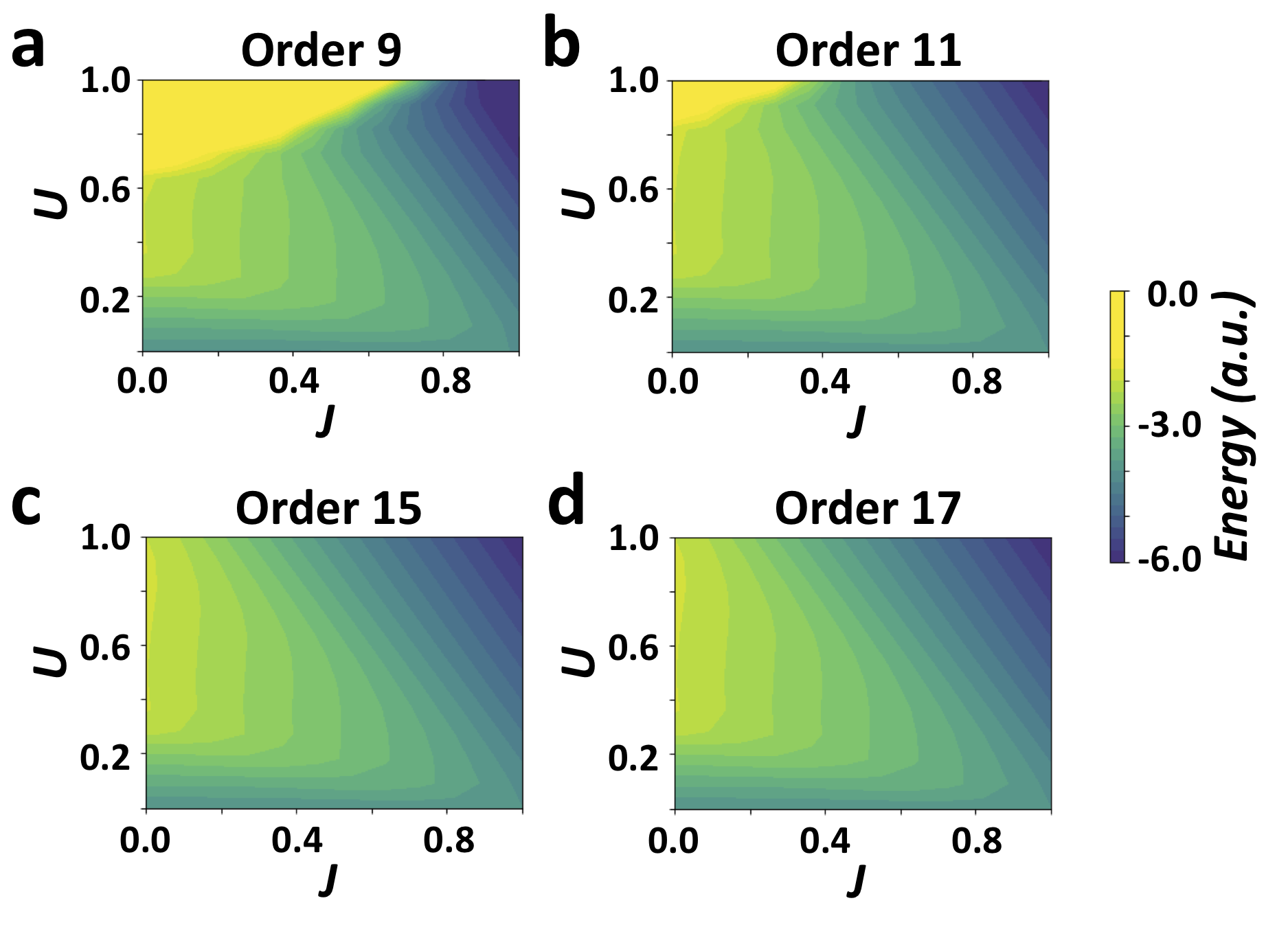}
\caption{Ground state energies with varied $U$'s and $J$'s estimated by the approximate 
ITE approaches employing (\textbf{a}) 9-th order, (\textbf{b}) 11-th order, 
(\textbf{c}) 15-th order, and (\textbf{d}) 17-th order Taylor expansion, respectively. 
For all the approximations, we fixed $\tau=2.5$.} \label{taylor}
\end{figure}

After fixing $\tau = \tau_{\rm opt}$, we can test the accuracy of the truncated Taylor expansion using different numbers of terms. As shown in Fig. \ref{taylor} the level of accuracy improves as more terms used in the truncated Taylor expansion, and it verifies that at least a 15th order expansion is needed in order to give accurate energy estimates for the entire $U$ and $J$ ranges. 
%

\section{Conclusion and outlook}

In this tutorial review, we have gone over some recent developments of hybrid quantum-classical algorithms based upon quantum 
computation of Hamiltonian moments. In particular, we have given an overview about how to rely on quantum resources to compute Hamiltonian 
moments, and what classical methods can be used following the quantum computation of the Hamiltonian moments to give rise to energy 
estimation for the target state. The levels of accuracy of these hybrid quantum-classical approaches are usually systematically improvable, 
even under the condition of working with noisy Hamiltonian moments, and thus provide error mitigation at the algorithmic level regardless of 
detailed description of the noise. Essentially, the algorithmic error mitigation can be attributed to the fact that these 
hybrid approaches are exploring the Krylov subspace of the entire Hilbert space, and if the ansatz is not totally orthogonal 
the true state, the power iteration would in principle guarantee a convergence given sufficiently high order Hamiltonian moments. 
On the other hand, what comes with the systematically improvable accuracy that is based on the cumulative construction of Krylov subspace is 
the price paid for obtaining higher order Hamiltonian moments, which will significantly increase the measurement requirement and bring 
hurdles for targeting larger quantum systems. The ongoing efforts are focused on either introducing some more efficient grouping rules 
to reduce the number of terms to be measured, or polishing the quality of the ansatz based on some physical and chemical inspiration 
such that lower order Hamiltonian moments will be sufficient for approaching certain level of accuracy. The latter can even be 
combined with variational quantum solver for the same purpose, which also helps the variational solver circumvent the barren plateaus. 
Another active direction is to combine these techniques with complete active space and/or Hamiltonian downfolding techniques 
(for recent studies, see e.g. Refs. \citenum{bauman2019downfolding,mniszewski2021reduction}) that can target larger systems. 
All these combinations are worth exploring in the NISQ era. Ultimately, from the accuracy and efficiency improvements brought 
by these algorithms and/or their combinations with some other techniques, the critical question becomes which problems are most 
benefited from quantum computed Hamiltonian moments and how much quantum advantage can be eventually gained in these contexts. \\

\section{Acknowledgement}
J. C. A., T. K., and B. P. acknowledge Dr. Karol Kowalski and Dr. Niri Govind for their helpful discussions on some topics during the preparation of this paper. 

\section{Funding Information}
J. C. A. was supported in part by the U.S. Department of Energy, Office of Science, Office of Workforce Development for Teachers and 
Scientists (WDTS) under the Science Undergraduate Laboratory Internships Program (SULI). 
T. K. is supported by the U. S. Department of Energy, Office of Science, Office of Workforce Development for Teachers and Scientists, 
Office of Science Graduate Student Research (SCGSR) program. The SCGSR program is administered by the Oak Ridge Institute for Science 
and Education (ORISE) for the DOE. ORISE is managed by ORAU under contract number DE-SC0014664. 
B. P. was supported by the 
“Embedding QC into Many-body Frameworks for Strongly Correlated Molecular and Materials Systems” project, which is 
funded by the U.S. Department of Energy, Office of Science, Office of Basic Energy Sciences (BES), the Division of Chemical Sciences, 
Geosciences, and Biosciences. This material is based upon work supported by the U.S. Department of Energy, Office of Science, National 
Quantum Information Science Research Centers. 

\section{Data Availability Statement}
The data and code that support the findings of this study are available at
\url{https://github.com/jaulicino/HamiltonianMoment/}

\bibliography{ref}
\end{document}